\documentclass[twocolumn,showpacs,prx,aps,longbibliography,superscriptaddress]{revtex4-1}
\usepackage[bookmarks=false,linkcolor=blue,urlcolor=blue,colorlinks,citecolor=blue]{hyperref}

\pdfoutput=1 

\usepackage{xcolor}

\usepackage{graphicx}
\usepackage{color}

\usepackage{lineno}

\begin{document}
\author{Maciej Koch-Janusz}
\affiliation{Institute for Theoretical Physics, ETH Zurich, 8093 Zurich, Switzerland}
\author{Zohar Ringel}
\affiliation{Racah Institute of Physics, The Hebrew University of Jerusalem, Jerusalem 9190401, Israel}

\title{Mutual Information,  Neural Networks and the Renormalization Group}

\begin{abstract}
Physical systems differring in their microscopic details often display strikingly similar behaviour when probed at macroscopic scales. Those universal properties, largely determining their physical characteristics, are revealed by the powerful renormalization group (RG) procedure, which systematically retains ``slow'' degrees of freedom and integrates out the rest. However, the important degrees of freedom may be difficult to identify. Here we demonstrate a machine learning  algorithm capable of identifying the relevant degrees of freedom and executing RG steps iteratively without any prior knowledge about the system. We introduce an artificial neural network  based on a model-independent, information-theoretic characterization of a real-space RG procedure, performing this task.  We apply the algorithm to classical statistical physics problems in one and two dimensions. We demonstrate RG flow and extract the Ising critical exponent. Our results demonstrate that machine learning techniques can extract abstract physical concepts and consequently become an integral part of theory- and model-building.

\end{abstract}

\maketitle

Machine learning  has been captivating public attention lately due to groundbreaking advances in automated translation, image and speach recognition \cite{lecun1},  game-playing \cite{silver}, and achieving super-human performance in tasks in which humans excelled while more traditional algorithmic approaches struggled \cite{Hershey:2010:SMS:1598088.1598466}. The applications of those techniques in physics are very recent, initially leveraging the trademark prowess of machine learning in classification and pattern recognition and applying them to classify phases of matter \cite{melko1,melko2,evert,wanglei,ohtsuki},  study amorphous materials \cite{Ronhovde2011,Ronhovde2012}, or exploiting the neural networks' potential as efficient non-linear approximators of arbitrary functions \cite{hinton1,tegmark} to introduce  a new numerical simulation method for quantum systems \cite{troyer,dassarma}. However, the exciting possibility of employing machine learning not as a numerical simulator, or a hypothesis tester, but as an integral part of the physical \emph{reasoning} process is still largely unexplored and, given the staggering pace of progress in the field of artificial intelligence, of fundamental importance and promise. 

 The renormalization group (RG) approach has been one of the conceptually most profound tools of theoretical physics since its inception. It underlies the seminal work on critical phenomena \cite{RevModPhys.47.773}, the discovery of asymptotic freedom in quantum chromodynamics \cite{PhysRevLett.30.1346}, and of the Kosterlitz-Thouless phase transition \cite{ber,kt}. The RG is not a monolith, but rather a conceptual framework comprising different techniques: real-space RG \cite{Kadanoff}, functional RG \cite{WETTERICH199390}, density matrix renormalization group (DMRG) \cite{PhysRevLett.69.2863}, among others. While all those schemes differ quite substantially in details, style and applicability there is an underlying physical intuition which encompasses all of them -- the essence of RG lies in identifying the ``relevant'' degrees of freedom and integrating out the ``irrelevant'' ones iteratively, thereby arriving at a universal, low-energy effective theory.  However potent the RG idea, those relevant degrees of freedom need to be identified first \cite{PhysRevLett.43.1434,PhysRevB.87.115144}. This is often a challenging conceptual step, particularly for strongly interacting systems and may involve a sequence of mathematical mappings to models, whose behaviour is better understood \cite{PhysRevB.87.161118,Auerbach}.
 
Here we introduce an artificial neural network algorithm iteratively identifying the physically relevant degrees of freedom in a spatial region and performing an RG coarse-graining step. The input data are samples of the system configurations drawn from a Boltzmann distribution, no further knowledge about the microscopic details of the system is provided. The internal parameters of the network, which ultimately encode the degrees of freedom of interest at each step, are optimized ('learned', in neural networks parlance) by a training algorithm based on evaluating real-space mutual information (RSMI) between spatially separated regions. We validate our approach by studying the Ising and dimer models of classical statistical physics in two dimensions. We obtain the RG flow and extract the Ising critical exponent. The robustness of the RSMI algorithm to physically irrelevant noise is demonstrated.

The identification of the important degrees of freedom, and ability to execute real-space RG procedure \cite{Kadanoff}, has significance not only quantitative but also conceptual: it allows to gain insights about the correct way of thinking about the problem at hand, raising the prospect that machine learning techniques may augment the scientific inquiry in a fundamental fashion.
.

\section*{The Real Space Mutual Information  algorithm}

Before going into more detail, let us provide a bird's eye view of our method and results. We begin by phrasing the problem in probabilistic/information-theoretic terms, a language also used in Refs. \cite{PhysRevD.54.5163,preskill,apenko,sethna,1367-2630-17-8-083005}. To this end, we consider a small ``visible'' spatial area $\mathcal{V}$, which together with its environment $\mathcal{E}$ forms the system $\mathcal{X}$, and we define a particular conditional probability distribution $P_{\Lambda}(\mathcal{H}|\mathcal{V})$, which describes how the relevant degrees of freedom $\mathcal{H}$  (dubbed ``hiddens'') in $\mathcal{V}$ depend on both $\mathcal{V}$ and $\mathcal{E}$. We then show that the sought-after conditional probability distribution is found by an algorithm maximizing an information-theoretic quantity, the mutual information (MI), and that this algorithm lends itself to a natural implementation using artificial neural networks. We describe how RG is practically performed by coarse-graining with respect to $P_{\Lambda}(\mathcal{H}|\mathcal{V})$ and iterating the procedure. Finally, we provide a verification of our claims by considering two paradigmatic models of statistical physics: the Ising model -- for which the RG procedure yields the famous Kadanoff block spins -- and the dimer model, whose relevant degrees of are much less trivial. We reconstruct the RG flow of the Ising model and extract the critical exponent.

Consider then a classical system of local degrees of freedom $\mathcal{X} = \{ x_1,\ldots, x_N \} \equiv \{x_i\}$, defined by a Hamiltonian energy function ${\rm {H}}(\{x_i\})$ and associated statistical probabilities $P(\mathcal{X}) \propto e^{-\beta {\rm H}(\{ x_i \})}$, where $\beta$ is the inverse temperature. Alternatively (and sufficiently for our purposes), the system is given by Monte Carlo (MC) samples of the equilibrium distribution $P(\mathcal{X})$. We denote a small spatial region of interest by $\mathcal{V} \equiv\{v_i\}$ and the remainder of the system by $\mathcal{E}\equiv \{e_i\}$, so that $\mathcal{X} = (\mathcal{V},\mathcal{E})$.  We adopt a probabilistic point of view, and treat $\mathcal{X}, \mathcal{E} $ etc. as random variables. Our goal is to extract the relevant degrees of freedom $\mathcal{H}$ from $\mathcal{V}$.

``Relevance'' is understood here in the following way: the degrees of freedom RG captures govern the long distance behaviour of the theory, and therefore the experimentally measurable physical properties;  they carry the most information about the system at large, as opposed to local fluctuations.  
We thus formally define the random variable $\mathcal{H}$ as a composite function of degrees of freedom in $\mathcal{V}$ maximizing the \emph{mutual information} (MI)  between $\mathcal{H}$ and the environment $\mathcal{E}$.  This definition, as we discuss in the Supplementary Materials, is related to the requirement that the effective coarse-grained Hamiltonian be compact and short-ranged, which is a condition any succesful standard RG scheme should satisfy. As we also show, it is supported by numerical results.

Mutual information, denoted by $I_{\lambda}$, measures the \emph{total} amount of information about one random variable contained in the other \cite{PhysRevLett.112.127204,Ronhovde2011,Ronhovde2012} (thus it is more general than correlation coefficients). It is given in our setting by:
\begin{equation}\label{mi1}  I_{\Lambda}(\mathcal{H}:\mathcal{E}) = \sum_{\mathcal{H},\mathcal{E}} P_{\Lambda}(\mathcal{E},\mathcal{H})\log\left( \frac{P_{\Lambda}(\mathcal{E},\mathcal{H})}{P_{\Lambda}(\mathcal{H})P(\mathcal{E})} \right),  \end{equation}
The unknown distribution $P_{\Lambda}(\mathcal{E},\mathcal{H})$ and its marginalization $P_{\Lambda}(\mathcal{H})$, depending on a set of parameters $\Lambda$ (which we keep generic at this point), are functions of 
$P(\mathcal{V}, \mathcal{E})$ and of $P_{\Lambda}(\mathcal{H}|\mathcal{V})$, which is the central object of interest. 

\begin{figure}[h]
	\centering
	\includegraphics*[width=8cm]{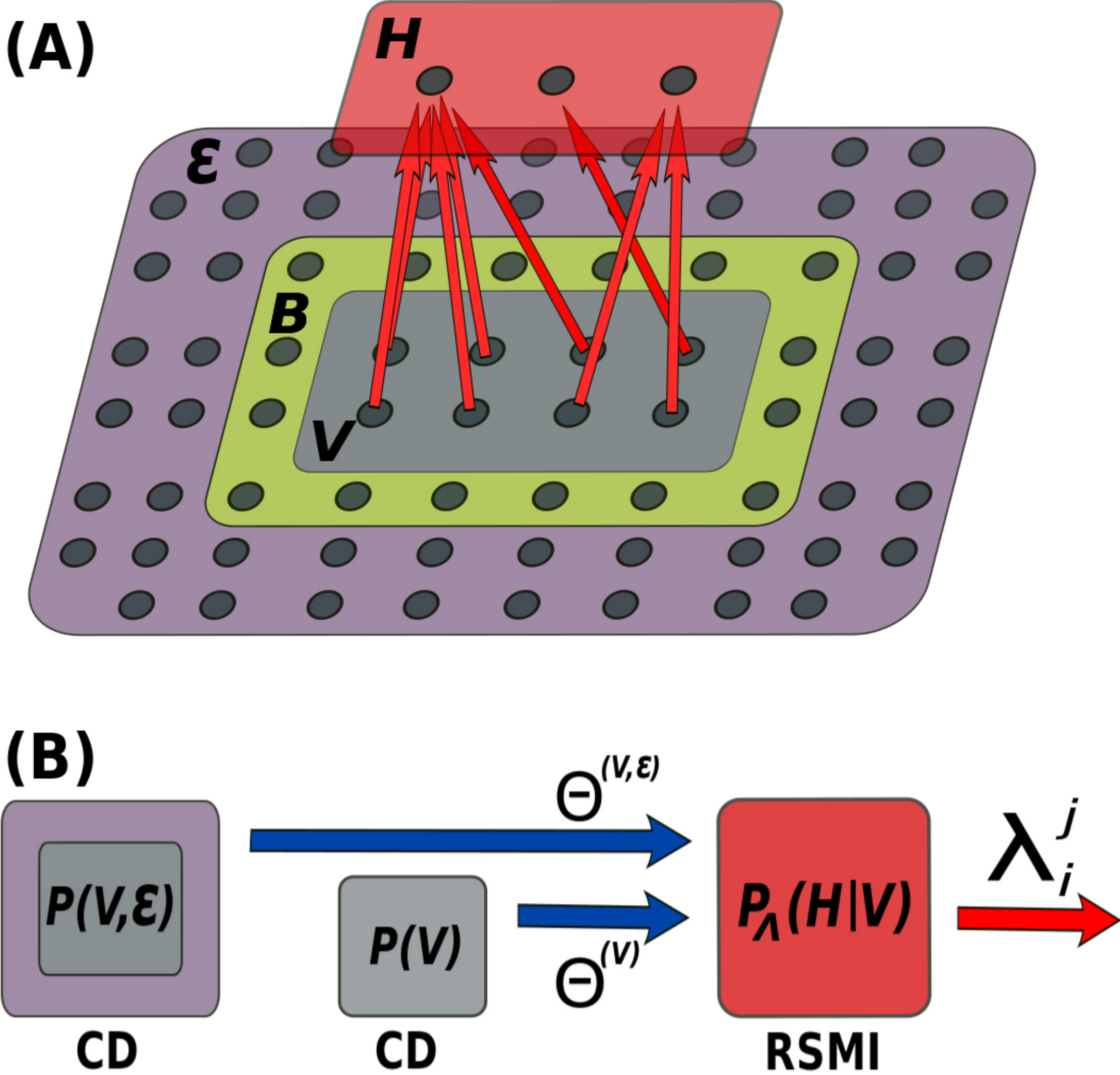}
	\caption{(A) The RSMI neural network architecture: the hidden layer $\mathcal{H}$ is directly coupled to the visible layer $\mathcal{V}$ via the  weights $\lambda^j_i$ (red arrows), however the training algorithm for the weights estimates MI between $\mathcal{H}$ and the environment $\mathcal{E}$. The buffer $\mathcal{B}$,  is introduced to filter out local correlations within $\mathcal{V}$ (see supplementary materials). (B) The workflow of the algorithm: the CD-algorithm trained RBMs learn to approximate probabilty distributions $P(\mathcal{V},\mathcal{E})$ and $P(\mathcal{V})$. Their final parameters, denoted collectively by $\Theta^{(\mathcal{V},\mathcal{E})}$ and $\Theta^{(\mathcal{V})}$, are inputs for the main RSMI network learning to extract $P_{\Lambda}(\mathcal{H}|\mathcal{V})$ by maximizing $I_{\Lambda}$. The final weights $\lambda^j_i$ of the RSMI network identify the relevant degrees of freedom. For Ising and dimer problems they are shown in Figs. 2 and 4.}
	\label{fig:rsmi}
\end{figure}

Finding $P_{\Lambda}(\mathcal{H}|\mathcal{V})$ which maximizes $I_{\Lambda}$ under certain constraints is a well-posed mathematical question and has a \emph{formal} solution \cite{infbottle1}. Since, however, the space of probability distributions grows exponentially with number of 
local degrees of freedom, it is in practice impossible to use without further assumptions for any but the smallest physical systems. Our approach is to exploit the remarkable dimensionality reduction properties of artificial neural networks \cite{hinton1}. We use restricted Boltzmann machines (RBM), a class of probabilistic networks well adapted to approximating arbitrary data probability distributions.
An RBM is composed of two layers of nodes, the ``visible'' layer, corresponding to local degrees of freedom in our setting, and a ``hidden'' layer.
The interactions between the layers are defined by an energy function $E_{\Theta}\equiv E_{a,b,\theta}(\mathcal{V},\mathcal{H}) = -\sum_i a_i v_i - \sum_j b_j h_j - \sum_{ij}v_i \theta_{ij}h_j$, such that the joint probability distribution for a particular configuration of visible and hidden deegrees of freedom is given by a Boltzmann weight:
\begin{equation}\label{mi3}  P_{\Theta}(\mathcal{V},\mathcal{H}) = \frac{1}{\mathcal{Z}}e^{-E_{a,b,\theta}(\mathcal{V},\mathcal{H})}, \end{equation}
with $\mathcal{Z}$ the normalization. The goal of the network training  is to find  parameters $\theta_{ij}$ (``weights'' or ``filters'') and $a_i,b_i$  optimizing a chosen objective function. 

Three distinct RBMs are used: two are trained as efficient approximators of the probability distributions  $P(\mathcal{V},\mathcal{E})$ and $P(\mathcal{V})$, using the celebrated contrastive divergence (CD) algorithm \cite{hinton2}. Their trained parameters are used by the third network [see Fig. 1(B)], which has a different objective:
to find $P_{\Lambda}(\mathcal{H}|\mathcal{V})$ maximizing $I_{\Lambda}$, we introduce the real space mutual information (RSMI) network, whose architecture is shown in Fig. 1(A). The hidden units of RSMI correspond to coarse-grained variables $\mathcal{H}$.

The parameters $\Lambda = (a_i,b_j,\lambda_i^j)$ of the RSMI network are trained by an iterative procedure. At each iteration a Monte Carlo estimate of function $I_{\Lambda}(\mathcal{H}:\mathcal{E})$ and its gradients is performed for the current values of parameters $\Lambda$. The gradients are then used to improve the values of weights in the next step, using a stochastic gradient descent procedure. 

The trained weights $\Lambda$ define the probability $P_{\Lambda}(\mathcal{H}|\mathcal{V})$  of a Boltzmann form, which is used to generate MC samples of the coarse-grained system. Those, in turn, become input to the next iteration of the RSMI algorithm. The  estimates of mutual information, weights of the trained RBMs and sets of generated MC samples at every RG step can be used to extract quantitative information about the system in the form of correlation functions, critical exponents etc. as we show below and in the supplementary materials. We also emphasize that the parameters $\Lambda$ identifying relevant degrees of freedom are re-computed at every RG step. This potentially allows RSMI to capture the evolution of the degrees of freedom along the RG flow \cite{LUDWIG1987687}.

\section*{Validation}

To validate our approach we consider two important classical models of statistical physics: the Ising model, whose coarse-grained degrees of freedom resemble the original ones, and the fully-packed dimer model, where they are entirely different. 

The Ising Hamiltonian on a two-dimensional square lattice  is:
\begin{equation}\label{ising} H_I = \sum_{<i,j>}s_is_j, \end{equation}
with $s_i = \pm1$ and the summation over nearest neighbours. Real-space RG of the Ising model proceeds by the block-spin construction \cite{Kadanoff}, whereby each $2\times 2$ block of spins is coarse grained into a single  effective spin, whose orientation is decided by a ``majority rule''.
\begin{figure}[h]
	\centering

	\includegraphics*[width=9cm]{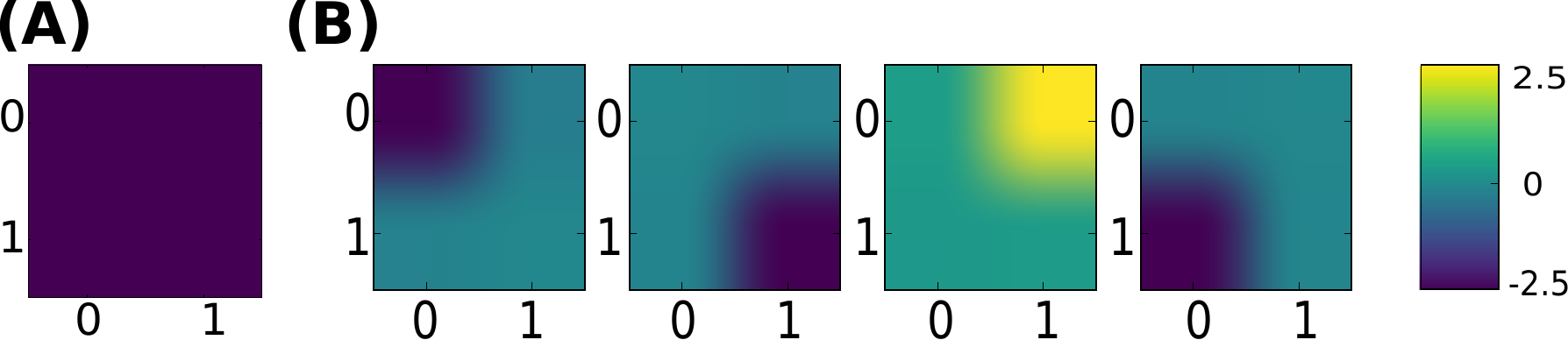}
	\caption{The weights of the RSMI network trained on Ising model. The ANN couples strongly to areas with  large absolute value of the weights. (A)$N_h=1$ hidden neuron, $\mathcal{V}$ area of $2 \times 2$ spins: the ANN discovers Kadanoff blocking (B) $N_h=4$ for $2 \times 2$ area $\mathcal{V}$. }
	\label{fig:ising1}
\end{figure}

The results of the RSMI algorithm trained on Ising model samples are shown in Fig. 2. We vary the number of both hidden neurons $N_h$ and the visible units, which are arranged in a 2D area $\mathcal{V}$ of size $L\times L$ [see Fig. 1(A)]. For a $4$ spin area the network indeed rediscovers the famous Kadanoff block-spin: Fig. 2(A) shows a single hidden unit coupling uniformly to $4$ visible spins, i.e. the orientation of the hidden unit is decided by the average magnetisation in the area. Fig. 2(B) is a trivial but important sanity check: given $4$ hidden units to extract relevant degrees of freedom from an area of $4$ spins, the networks couples each hidden unit to a different spin, as expected. In the supplementary materials we also compare the weights for areas $\mathcal{V}$ of different size, which are generalizations of Kadanoff procedure to larger blocks.

\begin{figure}[h]
	\centering
	\includegraphics*[width=9cm]{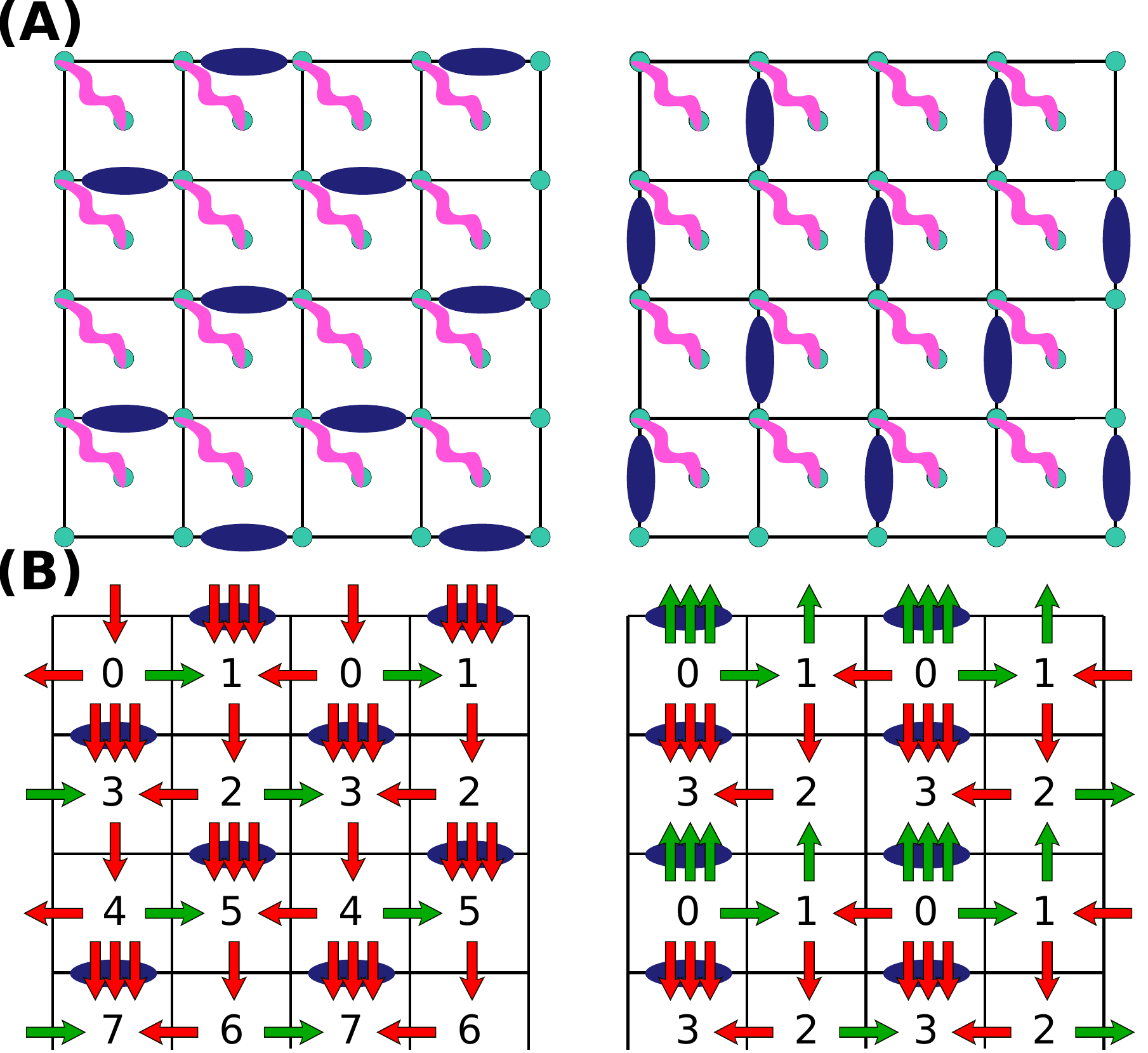}
	\caption{(A) Two sample dimer configurations (blue links), corresponding to $E_y$ and $E_x$ electrical fields, respectively. The coupled pairs of additional spin degrees of freedom on vartices and faces of the lattice (wiggly lines) are decoupled from the dimers and from each other. Their fluctuations constitute irrelevant noise. (B) An example of mapping the dimer model to local  electric fields. The so-called staggered configuration on the left maps to uniform nonvanishing field in the vertical direction: $\langle E_y\rangle \not= 0$. The ``columnar'' configuration on the right produces both $E_x$ and $E_y$ which are zero on average (see Ref. \cite{Fradkin} for details of the mapping).   }
	\label{fig:dimermapping}
\end{figure}
We next study the dimer model, given by an entropy-only partition function, which counts the number of dimer coverings of the lattice, i.e. subsets of edges such that every vertex is the endpoint of exactly one edge. Fig. 3(A) shows sample dimer configurations (and additional spin degrees of freedom added to generate noise). This deceptively simple description hides nontrivial physics \cite{PhysRev.132.1411} and correspondingly, the RG procedure for the dimer model is more subtle, since -- contrary to the Ising case -- the correct degrees of freedom to perform RG on are not dimers, but rather look like effective local electric fields. This is revealed by a mathematical mapping to a ``height field'' $h$ (see Figs.3(A,B) and Ref. \cite{Fradkin}), whose gradients behave like electric fields. The continuum limit of the dimer model is given by the following action:
\begin{equation}\label{dimfieldth} S_{dim}[h] = \int d^2x\ \left( \nabla h(\vec{x}) \right)^2 \equiv \int d^2x\  \vec{E}^2(\vec{x}), \end{equation}
and therefore the coarse-grained degrees of freedom are low-momentum (Fourier) components of the electrical fields $E_x, E_y$ in the $x$ and $y$ directions. They correspond to ``staggered'' dimer configurations shown in Fig. 3(A).
\begin{figure}[h]
	\centering
	\includegraphics*[width=8.8cm]{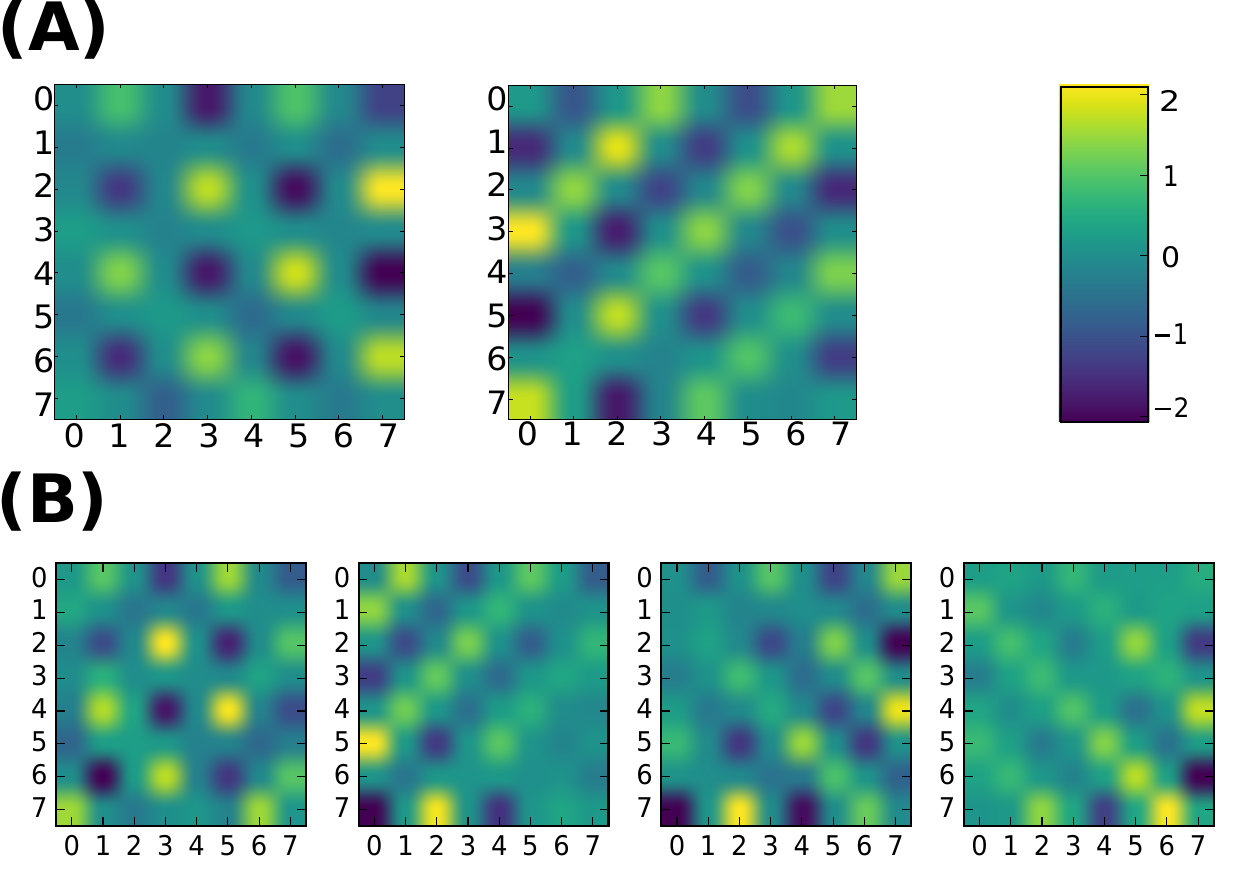}
	\caption{The weights of the RSMI network trained on dimer model data: (A) $N_h=2$ hidden neurons for a visible area $\mathcal{V}$ of $8\times 8$ spins. The two filters recognize $E_y$ and $E_x+E_y$ electrical fields, respectively [compare with dimer patterns in Fig. 3(A).].  (B) The trained weights for $N_h=4$ hidden neurons.}
	\label{fig:dimer1}
\end{figure}

Remarkably, the RSMI algorithm extracts the local electric fields from the dimer model samples without any knowledge of those mappings. In Fig. 4 the weights for $N_h=2$ and  $N_h=4$ hidden neurons, for an $8\times 8$ area [similar to Fig. 3(A)] are shown: the pattern of large negative (blue) weights couples strongly to a dimer pattern corresponding to local uniform $E_y$ field [see left pannels of Figs. 3(A,B)]. The large positive (yellow) weights select an identical pattern, translated by one link. The remaining neurons extract linear superpositions $E_x+E_y$ or $E_x-E_y$ of the fields.

To demonstrate the robustness of the RSMI, we added physically irrelevant noise, forming nevertheless a pronounced pattern, which we model by additional spin degrees of freedom, strongly coupled (ferromagnetically) in pairs [wiggly lines in Fig. 3(A)]. Decoupled from the dimers, and from other pairs, they form a trivial system, whose fluctuations are short-range noise on top of the dimer model. Vanishing weights [green in Figs. 4(A,B)] on sites where pairs of spins reside prove RSMI discards their fluctuations as irrelevant for long-range physics, despite their regular pattern. 

Notably, the filters obtained using our approach for the dimer model, which match the analytical expectation, are orthogonal to those obtained using Kullback-Leibler (KL) divergence. As expanded upon in the supplementary materials, this shows that standard RBMs minimizing the KL-divergence do not generally perform RG, thereby contradicting prior claims \cite{mehta}.

\begin{figure}[h]
	\centering
	\includegraphics*[width=8.5cm]{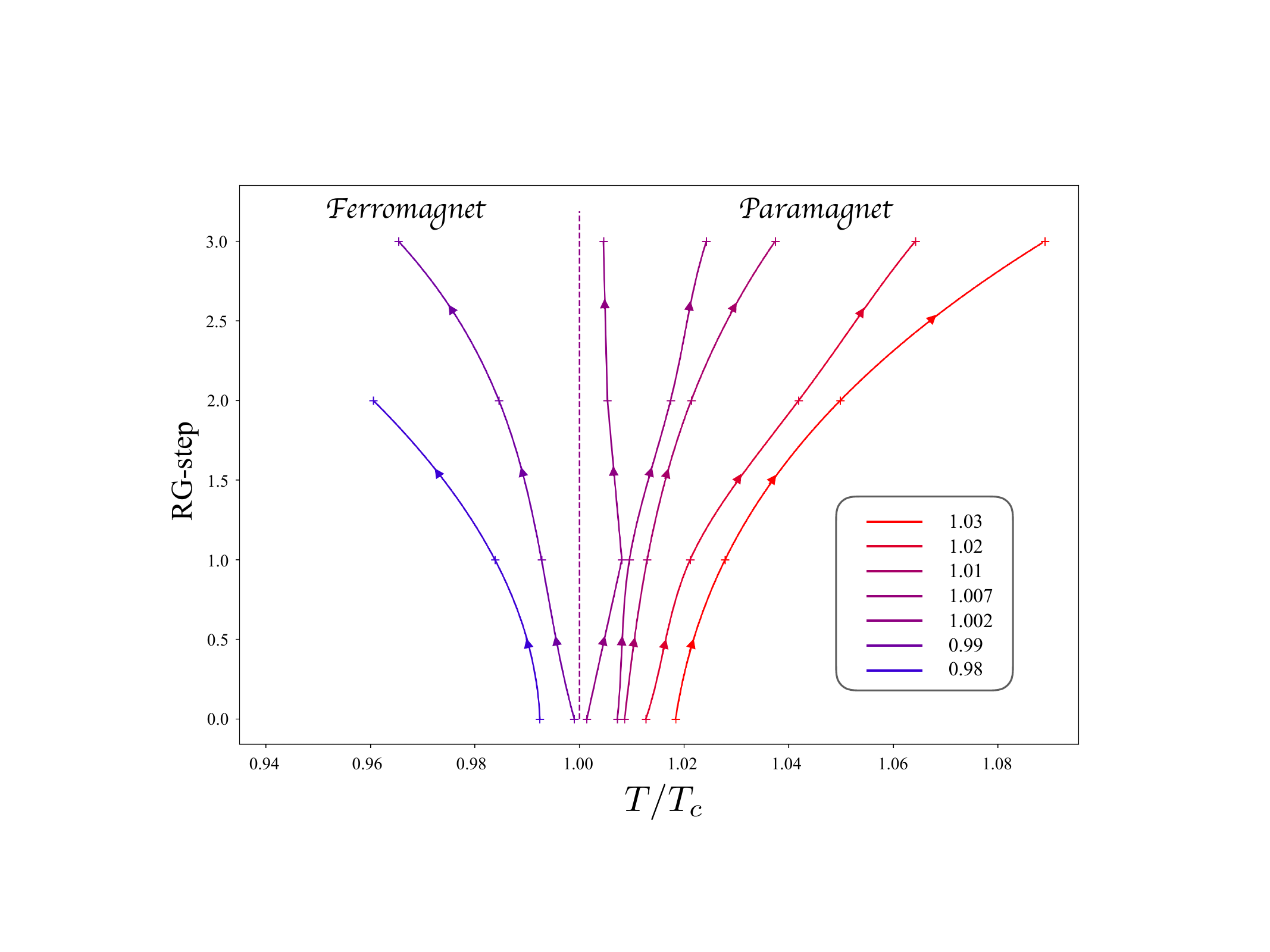}
	\caption{RG flow for the 2D Ising model.  The temperature $T$ (in units of $T_c$) as a function of the RG step for systems with initial temperatures below and above $T_c$.  }
	\label{fig:RG_flows}
\end{figure}

Finally, we demonstrate that by iterating the RSMI algorithm the qualitative insights about the nature of relevant degrees of freedom give rise to \emph{quantitive} results. To this end we revisit the 2D Ising model which (contrary to the dimer model) exhibits a nontrivial critical point at the temperature $T_c = (\log(1 + \sqrt{2})/2)^{-1}$, separating the paramagnetic and ferromagnetic phases. We generate MC samples of the system  of size $128\times 128$ at values $T$ around the critical point, and for each one we perform up to four RG steps, by computing the $\Lambda$ filters using RMSI, coarse-graining the system with respect to those filters (effectively halving the linear dimensions) and re-iterating the procedure. In addition to the set of MC configurations for the coarse-grained system, estimates of mutual information as well as the filters of the CD-trained RBMs, are generated and stored. The effective temperature $T$ of the system at each RG step can be evaluated entirely intrinsically either from correlations or  the mutual information, as discussed in the supplement. Using the RBM filters spin-spin correlations, next-nearest-neighbour, for instance, can be computed. By comparing these with known analytical results \cite{mccoy1973two} an additional cross-check of the effective temperature can be obtained.

In Fig. \ref{fig:RG_flows} the effective $T$ is plotted against $\log_2(\xi_{128} / \xi)$, where $\xi$, $\xi_{128}$  are the current and $128 \times 128$ system correlation lengths, respectively (this has the meaning of an RG step for integer values). The RG flow of the 2D Ising model is recovered: systems starting with $T < T_c$ flow towards ever decreasing $T$, i.e. an ordered state, while the ones with $T > T_c$ towards a paramagnet. In fact, the position of the critical point can be estimated with $1\%$ accuracy just from the divergent flow. Furthermore, we evaluate the correlation length exponent $\nu$, defined by $\xi \propto \tau^{-\nu}$. Using the finite-size data collapse [see Fig.4 in Supplemental Materials] its value, equal to the negative slope, is estimated to be $\nu \approx 1.0 \pm 0.15$ , consistent with the exact analytical result  $\nu = 1$

\section*{Future directions}
Artificial neural networks based on real-space mutual information optimization have proven capable of extracting complex information about physically relevant degrees of freedom and using it to perform real-space RG procedure. The RSMI algorithm we propose  allows for the study of existence and location of critical points, and RG flow in their vicinity, as well as estimation of correlations functions, critical exponents etc. 
This approach is an example of a new paradigm in applying machine learning in physics: the internal data representations discovered by suitably designed algorithms are not just technical means to an end, but instead are a clear reflection of the underlying structure of the physical system (see also \cite{Schoenholz}). Thus, in spite of their ``black box'' reputation, the innards of such architectures may teach us fundemental lessons.
This raises the prospect of employing machine learning in science in a collaborative fashion, exploiting the machines' power to distill subtle information from vast data, and human creativity and background knowledge \cite{mlreview}.

Numerous further research directions can be pursued. Most directly, equilibrium systems with less understood relevant degrees of freedom -- \emph{e.g.} disordered and glassy systems -- can be investigated \cite{Ronhovde2011,Ronhovde2012}. The ability of RSMI algorithm to re-compute the relevant degrees of freedom at every RG step potentially allows to study their evolution along the (more complicated) RG flow \cite{LUDWIG1987687}. Furthermore, though we studied classical systems, the extension to the quantum domain is possible via the quantum-to-classical mapping  of Euclidean path integral formalism.   A more detailed analysis of the mutual-information based RG procedure may prove fruitful from theory perspective. Finally, applications of RSMI beyond physics are possible, since it offers a neural network implementation of a variant of the Information Bottleneck method \cite{infbottle1}, succesful in compression and clustering analyses \cite{Slonim:2000:DCU:345508.345578}; it can also be used as a local-noise filtering pre-training stage for other machine learning algorithms.

\paragraph*{Acknowledgements --} We thank Profs. S. Huber and P. Fendley for discussions. M.K-J. gratefully acknowledges the support of Swiss National Science Foundation (SNSF). Z.R. was supported by the European Unions Horizon 2020 research and innovation programme under the Marie Sklodowska-Curie grant agreement No. 657111. 
\paragraph*{Authorship --} Both authors (MKJ, ZR) contributed equally to this work.
\paragraph*{Data availability statement --} The data that support the plots within this paper and other findings of this study are available from the corresponding author upon request.

\bibliography{mi_nat_new}

\pagebreak
\widetext
\begin{center}
	\textbf{Supplemental Materials for ``Mutual Information,  Neural Networks and the Renormalization Group'' -- Methods}
\end{center}

\section*{Estimating Mutual Information}

Here we gather the notations and define formally the quantities appearing in the main text. We derive in detail the expression for the approximate mutual information measure, which is evaluated numerically by the RSMI algorithm. This measure is given in terms of a number of probability distributions, accessible via Monte Carlo samples and approximated by contrastive divergence (CD) trained RBMs, or directly defined by (different) RBMs \cite{Haykin}.

We consider a statistical system $\mathcal{X} = \{ x_1,\ldots, x_N \} \equiv \{x_i\}_{i=1}^N $ of classical Ising variables $x_i \in \{0,1\}$, which can equally well describe the presence or absence of a dimer. The system $\mathcal{X}$ is divided into a small ``visible'' area $\mathcal{V}$, an ``environment'' $\mathcal{E}$, and a ``buffer'' $\mathcal{B}$ separating the degrees of freedom in $\mathcal{V}$ and $\mathcal{E}$ spatially (for clarity of exposition we assumed $\mathcal{B}=\emptyset$ in the main text). We additionally consider a set of  ``hidden'' Ising variables $\mathcal{H} = \{h_1,\ldots,h_{N_h}\}$. For the main RSMI network described in the text, $\mathcal{H}$ has the interpretation of coarse-grained degrees of freedom extracted from $\mathcal{V}$.

We assume the data distribution $P(\mathcal{X}) = P(\mathcal{V},\mathcal{B},\mathcal{E})$ -- formally a Boltzmann equilibrium distribution defined by a Hamiltonian $\rm H(\{x_i\})$ -- is \emph{given} to us indirectly via random (Monte Carlo) samples of $(\mathcal{V},\mathcal{B},\mathcal{E})_i$. The distributions $P(\mathcal{V},\mathcal{E})$ and $P(\mathcal{V})$ are defined as marginalizations of $P(\mathcal{X})$. Performing the marginalizations explicitly is computationally costly and therefore it is much more efficient to  approximate $P(\mathcal{V},\mathcal{E})$ and $P(\mathcal{V})$  using two RBMs of the type defined in and above  Eq. (2), trained using the CD-algorithm \cite{hinton2} on the restrictions of $(\mathcal{V},\mathcal{B},\mathcal{E})_i$ samples.  The trained networks, with parameters $\Theta^{(\mathcal{V,E})}$ and $\Theta^{(\mathcal{V})}$ (which we refer to as $\Theta$-RBMs) define probability distributions $P_{\Theta}(\mathcal{V},\mathcal{E})$ and $P_{\Theta}(\mathcal{V})$, respectively [see also Fig. 1(B)]. From mathematical standpoint, contrastive divergence is based on minimizing a proxy to the Kullback-Leibler divergence between $P_{\Theta}(\mathcal{V},\mathcal{E})$ and $P_{\Theta}(\mathcal{V})$ and the data probability distributions $P(\mathcal{V},\mathcal{E})$ and $P(\mathcal{V})$, respectively, i.e. the training produces RBMs which model the data  well \cite{hinton2}.

The conditional probability  distribution $P_{\Lambda}(\mathcal{H}|\mathcal{V})$ is \emph{defined} by another RBM, denoted henceforth by $\Lambda$-RBM, with tunable parameters $\Lambda = (a_i,b_j,\lambda_i^j)$:
\begin{eqnarray}
P_{\Lambda}(\mathcal{H}|\mathcal{V}) &=& \frac{e^{-E_{\Lambda}(\mathcal{V},\mathcal{H})}}{\sum_{\mathcal{H}} e^{-E_{\Lambda}(\mathcal{V},\mathcal{H})}}, \\ \nonumber 
E_{\Lambda}(\mathcal{V},\mathcal{H}) &=& \sum_{ij} -v_i\lambda_{i}^j h_j - \sum_ia_i v_i -\sum_j b_j h_j
\end{eqnarray}
In contrast to $\Theta$-RBMs it will \emph{not} be trained using CD-algorithm, since its objective \emph{is not} to approximate the data probability distribution. Instead, the parameters $\Lambda$ will be chosen so as to maximize a measure of mutual information between $\mathcal{V}$ and $\mathcal{E}$. The reason for exclusion of a buffer $\mathcal{B}$, generally of linear extent comparable to $\mathcal{V}$, is that otherwise MI would take into account correlations of $\mathcal{V}$ with its immediate vicinity, which are equivalent with short-ranged correlations within $\mathcal{V}$ itself.  We now derive the MI expression explicitly.

Using $P(\mathcal{V},\mathcal{E})$ and $P_{\Lambda}(\mathcal{H}|\mathcal{V})$ we can define the joint probability distribution  $P_{\Lambda}(\mathcal{V},\mathcal{E},\mathcal{H}) = P(\mathcal{V},\mathcal{E})P_{\Lambda}(\mathcal{H}|\mathcal{V})$ and marginalize over $\mathcal{V}$ to obtain 
$P_{\Lambda}(\mathcal{E},\mathcal{H})$. We can then define the mutual information (MI) between $\mathcal{E}$ and $\mathcal{H}$ in the standard fashion: 
\begin{equation}
I_{\Lambda}(\mathcal{H}:\mathcal{E}) = \sum_{\mathcal{H},\mathcal{E}} P_{\Lambda}(\mathcal{E},\mathcal{H}) \log \left( \frac{P_{\Lambda}(\mathcal{E},\mathcal{H})}{P_{\Lambda}(\mathcal{H})P(\mathcal{E})}\right)
\end{equation}
The main task is to find the set of parameters $\Lambda$ which maximizes $I_{\Lambda}(\mathcal{H}:\mathcal{E})$ given the samples $(\mathcal{V},\mathcal{E})_i$.
Since $P(\mathcal{E})$ is not a function of $\Lambda$ one can optimize a simpler quantity:
\begin{equation}
A_{\Lambda}(\mathcal{H}:\mathcal{E}) = \sum_{\mathcal{H},\mathcal{E}} P_{\Lambda}(\mathcal{E},\mathcal{H}) \log \left( \frac{P_{\Lambda}(\mathcal{E},\mathcal{H})}{P_{\Lambda}(\mathcal{H})}\right)
\end{equation}
Using the $\Theta$-RBM approximations of the data probability distributions as well as the definition of the $P_{\Lambda}(\mathcal{H},\mathcal{E})$ one can further rewrite this as:
\begin{equation}
A_{\Lambda}(\mathcal{H}:\mathcal{E}) = \sum_{\mathcal{H},\mathcal{E}} P_{\Lambda}(\mathcal{E},\mathcal{H}) \log \left( \frac{\sum_{\mathcal{V}} P_{\Lambda}(\mathcal{V},\mathcal{H}) P_{\Theta}(\mathcal{V},\mathcal{E})/P_{\Lambda}(\mathcal{V})}{\sum_{\mathcal{V}'} P_{\Lambda}(\mathcal{V}',\mathcal{H}) P_{\Theta}(\mathcal{V}')/P_{\Lambda}(\mathcal{V}')} \right)
\end{equation}
The daunting looking argument of the logarithm can in fact be cast in a simple form, using the fact that all the probability distributions involved either are of Boltzmann form, or marginalization thereof over the hidden variables, which can be performed explicitly:
\begin{equation}
A_{\Lambda}(\mathcal{H}:\mathcal{E}) \equiv \sum_{\mathcal{H},\mathcal{E}} P_{\Lambda}(\mathcal{E},\mathcal{H})\log \left( \frac{\sum_{\mathcal{V}} e^{-E_{\Lambda,\Theta}(\mathcal{V},\mathcal{E},\mathcal{H})}}{\sum_{\mathcal{V}'} e^{-E_{\Lambda,\Theta}(\mathcal{V}',\mathcal{H})}}\right),
\end{equation}
where 
\begin{eqnarray}
E_{\Lambda,\Theta}(\mathcal{V},\mathcal{E},\mathcal{H}) &=& E_{\Lambda}(\mathcal{V},\mathcal{H}) + E_{\Theta}(\mathcal{V},\mathcal{E}) + \sum_j \log[1+ \exp(\sum_i v_j \lambda_i^j + b_j)] \\ \nonumber 
E_{\Lambda,\Theta}(\mathcal{V},\mathcal{H}) &=& E_{\Lambda}(\mathcal{V},\mathcal{H}) + E_{\Theta}(\mathcal{V}) + \sum_j \log[1+ \exp(\sum_i v_j \lambda_i^j + b_j)], \\ \nonumber
\end{eqnarray}
and where  $E_{\Theta}(\mathcal{V},\mathcal{E})$ and $E_{\Theta}(\mathcal{V})$ are defined by the parameter sets $\Theta^{(\mathcal{V},\mathcal{E})}$ and $\Theta^{(\mathcal{V})}$ of the trained $\Theta$-RBMs:
\begin{eqnarray}
P_{\Theta}(\mathcal{V}) &\propto& e^{-E_{\Theta}(\mathcal{V})} \\ \nonumber
P_{\Theta}(\mathcal{V},\mathcal{E}) &\propto& e^{-E_{\Theta}(\mathcal{V},\mathcal{E})}
\end{eqnarray}
Note, that since in $P_{\Lambda}(\mathcal{H}|\mathcal{V})$ the $a_i$ parameter dependence cancels out [and consequently also in $P_{\Lambda}(\mathcal{E},\mathcal{H})$], the quantity $A_{\Lambda}$ does not depend on $a_i$. Hence, without loss of generality, we put $a_i \equiv 0$ in our numerical simulations, i.e. the $\Lambda$-RBM is specified by the set of parameters $\Lambda = (b_j, \lambda_i^j)$ only.

$A_{\Lambda}$ is an average over the distribution $P_{\Lambda}(\mathcal{E},\mathcal{H})$ of a logarithmic expression [see Eq. (5)], which itself can be further rewritten as a statistical expectation value for a system with energy $E_{\Lambda,\Theta}(\mathcal{V},\mathcal{H})$, with variables $\mathcal{H}$ held fixed:
\begin{eqnarray}
\log \left( \frac{\sum_{\mathcal{V}} e^{-E_{\Lambda,\Theta}(\mathcal{V},\mathcal{E},\mathcal{H})}}{\sum_{\mathcal{V}'} e^{-E_{\Lambda,\Theta}(\mathcal{V}',\mathcal{H})}}\right) &=& \log \left( \frac{\sum_{\mathcal{V}} e^{-E_{\Lambda,\Theta}(\mathcal{V},\mathcal{H}) - \Delta E_{\Lambda,\Theta}(\mathcal{V},\mathcal{E},\mathcal{H})}}{\sum_{\mathcal{V}'} e^{-E_{\Lambda,\Theta}(\mathcal{V}',\mathcal{H})}}\right) \\ 
&\equiv& \log \left( \left\langle e^{ - \Delta E_{\Lambda,\Theta}(\mathcal{V},\mathcal{E},\mathcal{H})}\right\rangle_{\mathcal{H}} \right) \approx \left\langle - \Delta E_{\Lambda,\Theta}(\mathcal{V},\mathcal{E},\mathcal{H})\right\rangle_{\mathcal{H}}  
\end{eqnarray}
with $\Delta E_{\Lambda,\Theta}(\mathcal{V},\mathcal{E},\mathcal{H}) =  E_{\Lambda,\Theta}(\mathcal{V},\mathcal{E},\mathcal{H})-E_{\Lambda,\Theta}(\mathcal{V},\mathcal{H})$.
Thus finally, we arrive at a simple expression for $A_{\Lambda}$:
\begin{equation} A_{\Lambda}(\mathcal{H}:\mathcal{E}) \approx \sum_{\mathcal{H},\mathcal{E}} P_{\Lambda}(\mathcal{E},\mathcal{H}) \left\langle - \Delta E_{\Lambda,\Theta}(\mathcal{V},\mathcal{E},\mathcal{H})\right\rangle_{\mathcal{H}}. \end{equation}
This expression can be numerically evaluated: using the fact that $P_{\Lambda}(\mathcal{E},\mathcal{H}) = \sum_{\mathcal{V}'}P_{\Lambda}(\mathcal{H}|\mathcal{V}')P(\mathcal{V}',\mathcal{E})$ we replace the sums over $\mathcal{V}'$ and $\mathcal{E}$ with a Monte Carlo (MC) average over $N_{(\mathcal{V},\mathcal{E})}$ samples $(\mathcal{V}',\mathcal{E})_i$. Furthermore, given a $\Lambda$-RBM (at current stage of training) and a sample of $(\mathcal{V})_i$, one can easily draw a sample $(\mathcal{H})_i \equiv (\mathcal{H}(\mathcal{V}))_i$ according to probability distribution $P_{\Lambda}(\mathcal{H}|\mathcal{V})$. Hence we have a MC estimate:
\begin{equation} \label{alambdafinal}A_{\Lambda}(\mathcal{H}:\mathcal{E}) \approx \frac{1}{N_{(\mathcal{V},\mathcal{E})}} \sum_{(\mathcal{V}',\mathcal{E},\mathcal{H}(\mathcal{V}'))_i} \left\langle - \Delta E_{\Lambda,\Theta}(\mathcal{V},\mathcal{E},\mathcal{H})\right\rangle_{\mathcal{H}}. \end{equation}
The expectation value in the summand is itself also evaluated by MC averaging, this time with respect to Boltzmann probability distribution with energy $E_{\Lambda,\Theta}(\mathcal{V},\mathcal{H})$.

\section*{Numerical evaluation}

\subsection{One step of RSMI}

\begin{figure}[h]
	\centering
	\includegraphics*[width=11cm]{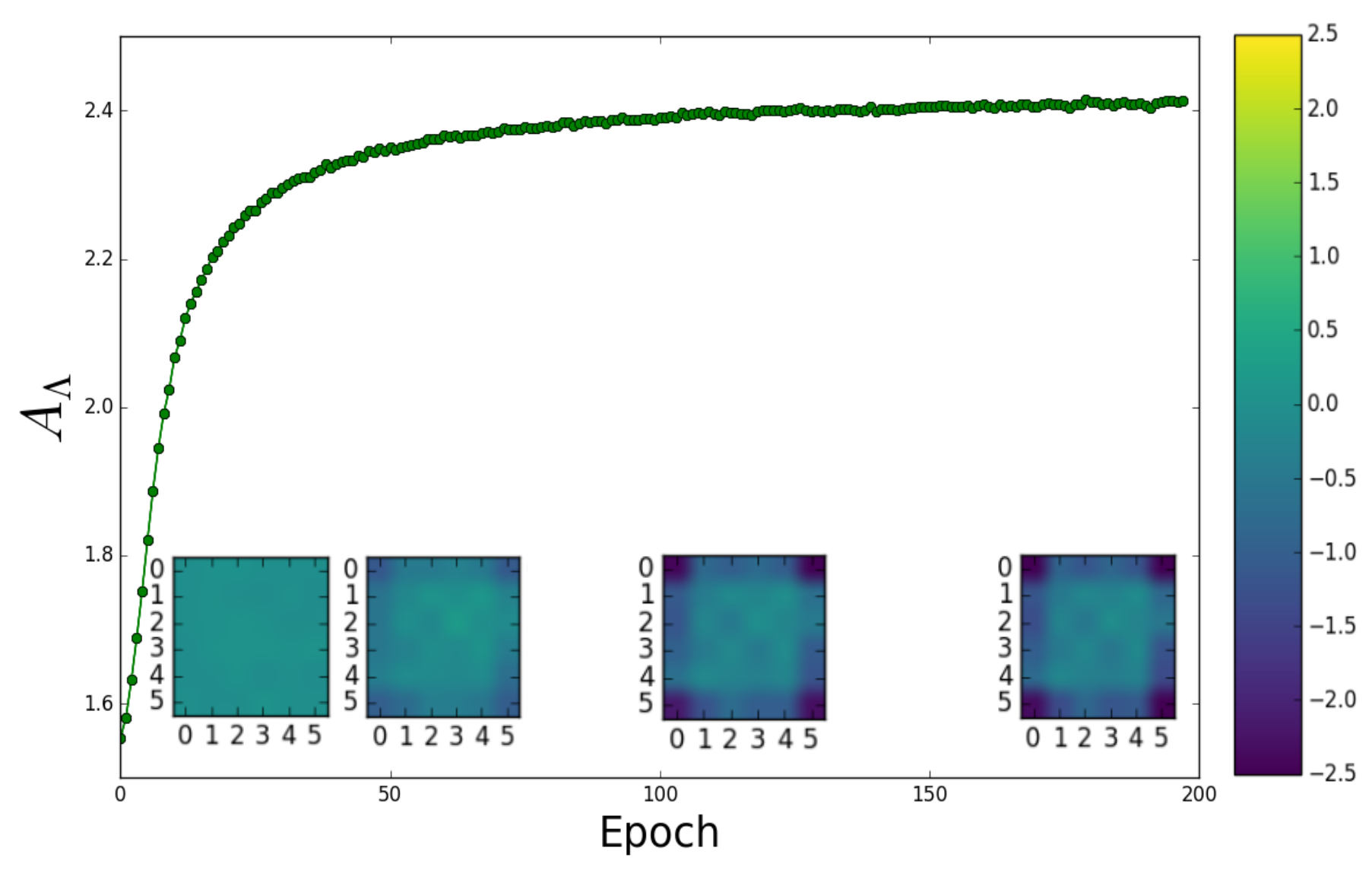}
	\caption{The proxy $A_{\Lambda}$ of mutual information, as a function of training epochs of the $\Lambda$-RBM for the Ising model data. The saturation behaviour of $A_{\Lambda}$ demonstrates the convergence of the algorithm. The weight matrices shown are for a single hidden neuron, with visible area $\mathcal{V}$ of size $6\times6$, after 0, 10, 100 and 190 training epochs respectively. The development of the boundary coupling behaviour for the 2D sing system, discussed in the supplemental materials, can be seen.}  
	\label{fig:ising2}
\end{figure}
Numerical results in the paper were obtained using a purpose-written (Python/Theano \cite{2016arXiv160502688short}) implementation of the main RSMI network and a standard implementation of CD-trained RBMs. For the RSMI network, maximizing $A_{\Lambda}$ with respect to parameters $\Lambda = (a_i,b_j,\lambda_i^j)$ is performed using stochastic gradient descent procedure \cite{Haykin}. To this end we estimate the derivative  $\partial_{\lambda_{ij}} A_{\Lambda}$ over samples $(\mathcal{V},\mathcal{E},\mathcal{H}(\mathcal{V}))_i$. More accurately, we divide the samples into mini-batches, obtain an average assessment of $\partial_{\lambda} A_{\Lambda}$, and use it to update the parameters of the $\Lambda$-RBM: $\lambda_i^j \rightarrow \lambda_i^j  -\eta\cdot \partial_{\lambda_{ij}} A_{\Lambda} $ (and similarly for $b_j$) with a learning rate $\eta$. This is then repeated for next mini-batch. A run through all mini-batches constitutes one epoch of training. For the final data from the RSMI network 2000 epochs were used with a mini-batch size of 800 with a learning rate of $\eta=0.01 - 0.05$;  a regulator was used. The internal Monte Carlo rough estimate of the expectation value in Eq. \ref{alambdafinal} used two samples after a burn-in period of 126. For numerical efficiency we restricted the size of the environment-buffer-visible area setup to a window of total size three times the linear extent of the visible area to be coarse-grained. The contrastive divergence RBMs were trained for 300 epochs in the case of Ising and 2000 epochs for the dimers with a mini-batch size of 25 and learning rate $\eta=0.025$;  a regulator was used.
Fig. 5  shows  convergence of the $A_{\Lambda}$ estimation and the development of the weight matrices for an example training run in the case of the Ising system.

The initial Ising/dimer data were generated by Monte Carlo simulations. For the dimer data we used a 64$\times 64$ lattice (128$\times 128$ with spins) and performed MC with loop updates. Number of steps $t_0$ was tuned to have sample-at-t and sample-at-t+$t_0$ correlation at the level of noise. For the Ising system we used 128$\times 128$  lattice and a cluster-update MC.

We remark that the gradients of $A_{\Lambda}$ should best be computed explicitly (a simple, if tedious, computation)
prior to numerical evaluation, and one should not use the automated gradient computation capability provided by \emph{e.g.} Theano package \cite{2016arXiv160502688short}. The reason is that some of the dependence on parameters $\Lambda$ is stochastic. Specifically, it enters via the treshold values of the Monte Carlo acceptance, and this dependence results in a piece-wise constant function (although with very fine steps) which is not handled correctly by automated gradient computing procedures (the numerical gradients would be equal to zero in most cases).

\subsection{Multiple steps}

\begin{figure}[h]
	\centering
	\includegraphics*[width=14cm]{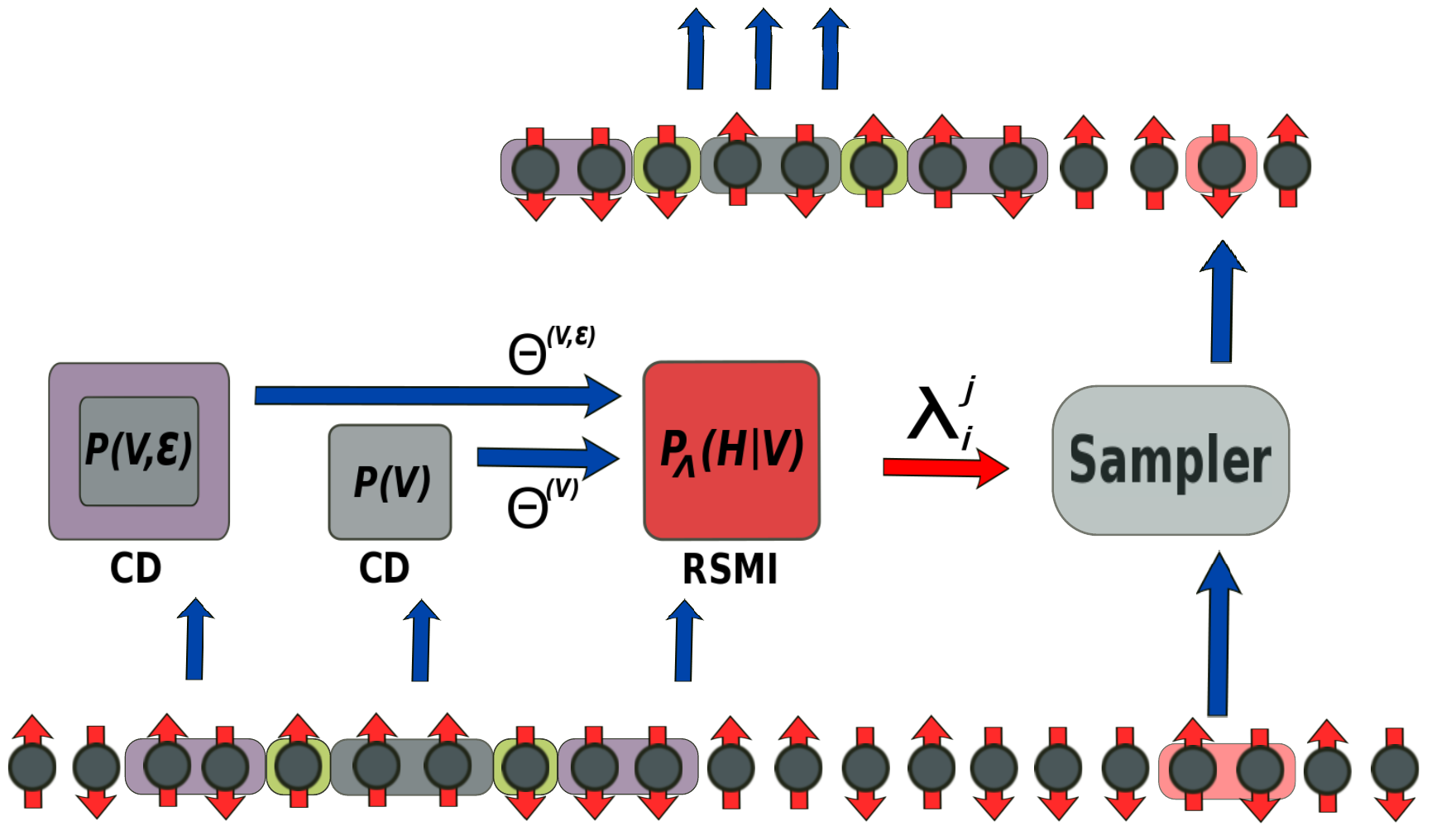}
	\caption{Data flow in multiple steps of RSMI algorithm: the configurations of the system of size $L$ (drawn in 1D for simplicity), are used to train the $\Theta$-RBMs, whose parameters are fed to the $\Lambda$-RBM. The samples are then used for training the $\Lambda$-RBM. Its trained filters, are fed, in turn, to a simple Monte Carlo sampler, which coarse-grains from the $L$-sized samples to generate new set of configurations of size $L/2$ (we take, for concreteness, a visible area of size 2, denoted by grey box surrounded by buffer and environment in the training stage, and by a red box encompassing two spins to be coarse-grained into a single one in the sampling stage). }  
	\label{fig:multisteps}
\end{figure}

Here we describe the iterative procedure for performing multiple RG steps in sequence and extracting the numerical quantities characterizing the RG flow. The structure of the algorithm is shown in Fig. \ref{fig:multisteps}. The trained filters of the $\Lambda$-RBM are used to construct a Monte Carlo sampler: the size $L$ configurations are tiled with a window of size $\mathcal{V}$ and $\mathcal{H}$ new coarse-grained variables are assigned with acceptance given by $P_{\Lambda}(\mathcal{H}|\mathcal{V})$. For the 2D Ising model we used a single coarse-grained variable for a visible area of $2\times 2$. The new set of configurations of size $L/2$ is used to train the $\Theta$- and $\Lambda$-RBMs at this scale, iterating the procedure.

In addition to the $\Lambda$ filters, whose significance in identifying the relevant degrees of freedom we discussed in the main text, the multiple stages of the RSMI algorithm generate a wealth of other numerical data of physical importance:
\begin{itemize}
	\item The sets of MC configurations at successive length-scales. Those can be used to compute correlations functions, evaluate expectation values etc.
	\item The $\Theta$-RBM filters at successive length-scales. Since the trained $\Theta$-RBM is an efficient approximator to the specific Boltzmann probability generating the samples of the system at a given length-scale, it can also be used to compute the properties of the system, such as correlations (without using the coarse-grained MC samples, in fact the RBM can be used to \emph{generate} new samples). Below we compute the next-nearest-neigbour correlations in the Ising model as an example. They can also be used to intrinsically evaluate the effective temperature of the system.
	\item The Mutual Information $I_\Lambda$ (or the proxy $A_{\Lambda}$) captured by the $\Lambda$-RBM. It can also be used to evaluate the effective temperature at succesive RG steps intrinsically. Below we show how how a practical MI ``thermometer'' is constructed.
	
	These data allows the RSMI algorithm to make quantitive predictions. We show below, on the example of 2D Ising model, how they are sufficient to characterize the RG flow: the position of the critical point, the flow around it (stable/unstable) as well as critical exponents can all be evaluated.
	
\end{itemize}  

We remark here that using coarse-graining schemes with more than one hidden degree of freedom per visible area requires certain care. This is well illustrated by Fig. 2B in the main text, where one of the hidden spins $h$ couples anti-ferromagnetically to 
the visible spin it is tracking.  The immediate reason is that MI is maximized when $h$ and $v$ are \emph{either} perfectly correlated \emph{or} perfectly anti-correlated, as in both cases knowing one spin of the pair fully determines the other. Thus, for every hidden,  there is a local $Z_2$ symmetry, which is generically not broken, since RBMs have the conditional independence property, i.e. $P(h_j,h_j| \mathcal{V}) = P(h_i|\mathcal{V})P(h_j|\mathcal{V})$. Thus every one of 4 hiddens independently ``decides" whether to align of anti-align with the single visible spin it is tracking. 

One may wonder if this is not a problem from the point of view of using the filters to generate coarse-grained configurations (for which the ``sanity check'' filters were never intended to be used; their only purpose is to demonstrate full information capture), as independent ``misaligned'' filters could, for instance, introduce anti-correlations between coarse-grained areas, where previously the spins were correlated, or vice versa.  The answer is, that this pitfall is in practice avoidable. First, in the most common case of coarse-graining an area to a single hidden degree of freedom there is only one filter, which is used for the whole system. For a coarse-graining with more hiddens the relative phase between the filters needs to be fixed, which can done, for instance,  by comparing the signs of the correlator $\langle h_jh_j\rangle$ computed using the coarse-grained variables to the correlator of the original (composite) variables the $h$ couple to, which is fast  for small areas $\mathcal{V}$.

\section*{RG flow}

In this subsection we provide more details  on how the RG flow results in the paper were obtained. We demonstrate the ability of the RSMI algorithm to characterize the RG flow both qualitatively and quantitatively on the canonical example of the 2D Ising model. This allows us to benchmark against exact analytical results for the Ising model. The model exhibits two phases: a high-temperature disordered (paramagnetic) phase, and a low-temperature ordered (ferromagnetic) phase, separated by a critical point, which is an unstable RG fixed point. The critical exponents are known exactly.

\begin{figure}[h]
	\centering
	\includegraphics*[width=13cm]{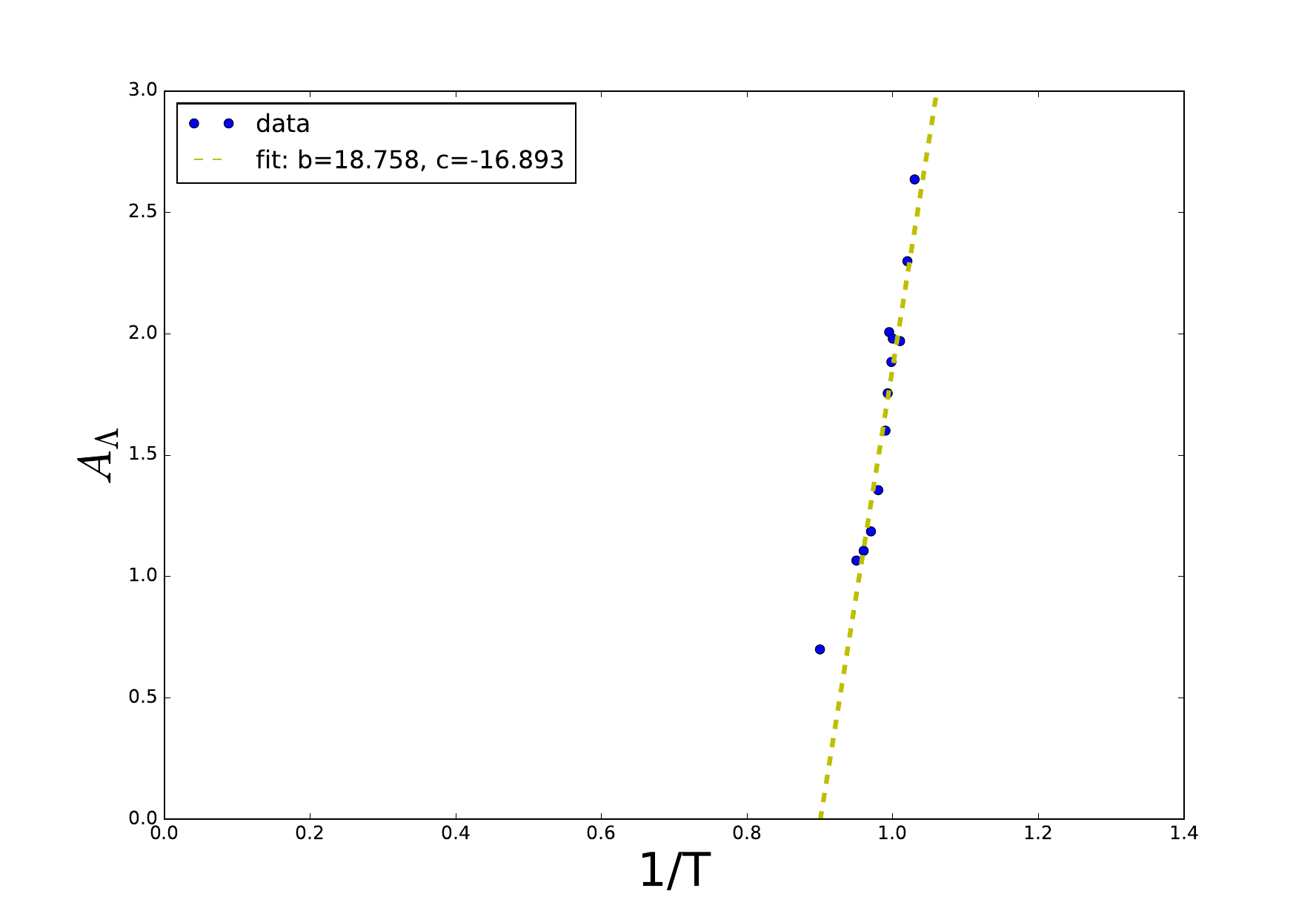}
	\caption{The dependence of mutual information proxy $A_{\Lambda}$ on the inverse temperature $\beta$ (in units of $\beta_{c}$), as extracted from the trained $\Lambda$-RBM for the $128\times 128$ Ising system. Within the range  $\beta \in [0.95,1.03]$ we fit a linear function $A_{\Lambda}(\beta) = b\cdot \beta + c$ to be used as a thermometric curve. The value of $A_\Lambda$ at the point $\beta = 0.9$ shows the expected deviation from linear behaviour (see discussion in the text). }  
	\label{fig:tempfitplot}
\end{figure}

In order to test our predictions we generated sets of MC samples of the system at values of the inverse temperature $\beta \in [0.95,1.03]$    (in units of the critical $\beta_{c}$). For each of the sets we first performed a single RG step i.e. we trained the $\Theta$- and $\Lambda$-RBMs. The next task is to establish a measure of the effective temperature of the system, or a ``thermometer''.  This is important,  as it can be used to estimate the effective temperature in subsequent RG steps completely intrinsically, i.e. without any need to rely on the knowledge of temperature dependence of some quantities obtained by other analytical or numerical methods (for a generic model, such knowledge may simply not be available).
This can be done in a number of ways. As we mentioned in the previus section, the $\Theta$-RBMs can be used to compute the spin-spin correlations. Since the correlations in the system have a clear temperature dependence, we can use the samples in the first step to establish an empirical curve $f(\beta)$, where $f$ is the chosen measure of correlations. In subsequent steps of the RG procedure $f$ can be evaluated either from the MC samples or the $\Theta$-RBMs and the effective temperature can be obtained as $\beta(f)$.
Alternatively, the value of mutual information $I_{\Lambda}$ (or, more precisely, of the proxy $A_{\Lambda}$) captured by the trained $\Lambda$-RBM can be used for this purpose. At $\beta = 0$ this value should be identically zero -- the system is totally uncorrelated -- while at $\beta \rightarrow \infty$ it is bounded from above by the total entropy of the variables in the visible area. In general, MI is a monotonic function of $\beta$, since any form of correlations decrease with temperature. Both methods give similar results.
In Fig. \ref{fig:tempfitplot} we plot the value of $A_{\Lambda}$  as a function of $\beta$ for the 2D Ising model.  In the parameter regime we considered we  fit a linear relation $A_{\Lambda}(\beta)$. The validity of using this simple fit is, of course, restricted to the range of values of $A_{\Lambda}$ for which it was performed. In general, as discussed above, the dependence in not linear, but a non-linear  fit can be used  in exactly the same way.

Having constructed the thermometer we are now equiped to perform subsequent RG steps, as described in the previous section. In Fig. 5 (in the main text) we show the effective temperature $T$ of the system as a function of the RG step, for initial value of $\beta \in [0.97, 1.02]$. We performed up to four RG steps (from $128\times 128$ system down to $16\times 16$). There are number of important observations to be made:
\begin{itemize}
	\item The effective $\beta$ of systems with initial $\beta < \beta_{c}$ consistently decreases with successive RG steps, while it increases for initial $\beta > \beta_{c}$.  For initial $\beta = \beta_{c}$ we see the subsequent values of $\beta$ are constant within accuracy. This is a signature of a divergent flow around the unstable  point $\beta = \beta_{c}$. 
	
	\item Examining the directions of the flow allows to establish the existence and nature (stable/unstable), and to find the position of the critical point. In fact, with the numerical data four our system,  within  $1\%$ accuracy.
	
	\item The flow data can be used to extract the value of the critical exponents (see discussion below)

\end{itemize}

\begin{figure}[h]
	\centering
	\includegraphics*[width=10cm]{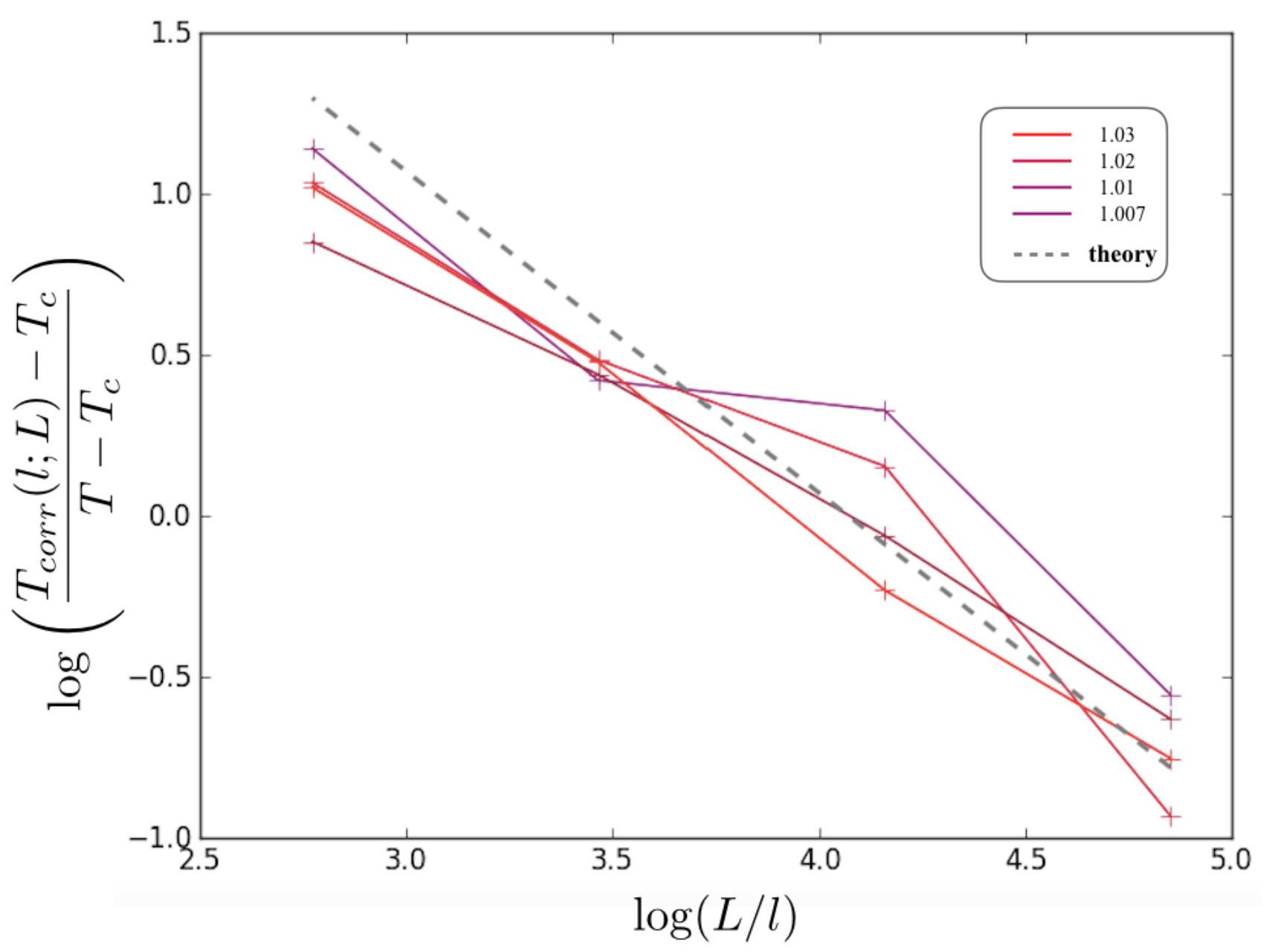}
	\caption{The finite size data collapse curves for the 2D Ising model (see discussion below the figure). The critical exponent consistent with the exact result $\nu=1$ can be read of from the slope.}  
	\label{fig:flowfig002}
\end{figure}

The two "thermometers" we used, based on mutual information and correlations, provide us with two inverse-temperature estimates $\beta_{MI}(l,\beta)$ and $\beta_{Corr}(l,\beta)$ which depend on the scale $l$ on which the measurement was carried and the microscopic inverse temperature $\beta$. In an infinite system these scale-dependent estimates will reflect the renormalisation group flow of the temperature, namely $(\beta_{*}(l)-\beta_c) = (\beta_{*}(l=a)-\beta_c)(l/a)^{1/\nu}$, with $\beta_{*}$ being the inverse temperature and $* \in \{MI, Corr\}$. In a finite system this behaviour is augmented by some scaling function $f_0(L,l,a)$ such that $(\beta_{*}(l;L)-\beta_c) = (\beta-\beta_c)(l/a)^{1/\nu} f_0(L,l,a)$, with $f_0$ to be determined and $a$ being the lattice spacing. The finite size scaling hypothesis (see for instance \cite{cardy1988finite}) allows us to simplify $f_0$ by taking $f_0(L,l,a) = f_0(L/l)$. We next define $\tilde{f}_0(x) = x^{-1} f_0(x^{\nu})$, and obtain 
\begin{equation}
\frac{\beta_*(l;L)-\beta_c}{\beta-\beta_c} = \tilde{f}_0\left( (L/l)^{1/\nu} \right).
\end{equation}

Following the normalization prescription, all the data point ($\beta_*(l;L)$) obtained for different values of $\beta$ and $l$, should collapse on a single curve defined by $\tilde{f}_0$.  The result of this exercise is shown in Fig. \ref{fig:flowfig002}, where we plot the data for four RG steps (for initial $\beta \in \{0.97,0.98,0.99\}$, i.e. the paramagnetic side of the transition). The slope of the curves is consistent with the exact analytical result for the exponent, i.e. $\nu =1$.

We note here that the errors can be  reduced by increasing the number of the MC samples (5000 per a value of $\beta$ in our simulations) and especially by considering larger initial system size ($128\times 128$ is the limit under reasonable time constraints for our Python code).

\section*{Relation of Mutual Information RG procedure to conventional RG schemes}
Here we provide an intuitive theoretical argument elucidating the connection of our mutual information based approach to more standard treatments of real-space RG. Various information-theoretic approaches to RG were also advocated or investigated in Refs. \cite{PhysRevD.54.5163,preskill,apenko,sethna,1367-2630-17-8-083005}.  Before defining our explicit criteria for identifying relevant degrees of freedom, let us first briefly rephrase the conventional  RG procedure in probabilistic terms. 

Consider then a physical system, represented by a set of local degrees of freedom (or random variables) $\mathcal{X}=\{x_i\}$ and governed by a Hamiltonian energy function ${\rm {H}}(\{x_i\})$. The equlibrium probability distribution is of Boltzmann form: $P(\{x_i\}) \propto e^{-\beta {\rm H}(\{ x_i \})}$, with $\beta$ the inverse temperature. Next we consider a new and smaller set of degrees of freedom $\mathcal{H}=\{h_i\}$, i.e. the coarse-grained variables, whose dependece on $\mathcal{X}$ is given by a conditional probability $\Pi_i P_{\Lambda}(h_i | \mathcal{X})$, where $\Lambda$ are variational internal parameters to be specified and each $h_i$ depends on some localized set of $\{x_i\}$. The RG transformation in this language consists of finding the effective Hamiltonian of the coarse-grained degrees of freedom ${\rm \tilde{H}}(\mathcal{H})$ by marginalizing over (or integrating-out, in physical terms) degrees of freedom $\mathcal{X}$ in the joint probability distribution of $\mathcal{X}$ and $\mathcal{H}$: 
\begin{eqnarray}
\label{Eq:expand}
{\rm \tilde{H}} &=& -\log(Z[\mathcal{H}]), \\ \nonumber 
Z[\mathcal{H}] &=& \sum_{\mathcal{X}} \Pi_i P_{\Lambda}(h_i | \mathcal{X}) e^{-\beta {\rm H}}.
\end{eqnarray}

Using the fact that all conditional probabilities are normalized we have that $\tilde{Z} \equiv \sum_{\mathcal{H}} Z[\mathcal{H}] = Z$, i.e. the new partition function  $\tilde{Z}$ has the exact same free energy as the original one $Z$. Notably it also contains all the information required to evaluate certain expectations values in an exact fashion. Consider for instance the average of $h_{j} h_k$ under $\tilde{Z}$. It can be re-expressed as an average over the original degrees of freedom as follows:
\begin{eqnarray}
\langle h_{j} h_k \rangle_{\tilde{Z}} &= \tilde{Z}^{-1}\sum_{\mathcal{H}} \sum_{\mathcal{X}} h_j h_k \Pi_i P_{\Lambda}(h_i | \mathcal{X}) e^{-\beta {\rm H}} \\ \nonumber
&= Z^{-1}\sum_{\mathcal{X}} [\sum_{h_j} h_jP_{\Lambda}(h_j | \mathcal{X})][\sum_{h_k} h_k P_{\Lambda}(h_k | \mathcal{X})] e^{-\beta {\rm H}}, 
\end{eqnarray}
where we used the fact that the overall conditional probability of $\mathcal{H}$ given $\mathcal{X}$ is given in terms of a product of $P(h_i | \mathcal{X})$, i.e. conditional independence, and in the second line we explicitly performed the trivial summations over $h_i \notin \{h_j,h_k\}$.  The summations over the remaining hiddens $h_j$ and $h_k$ can now be carried out,  yielding expectation values of $h_j$ and $h_k$ with  $\mathcal{X}$ held fixed, which we denote by $\bar{h}_j(\mathcal{X})$ and $\bar{h}_k(\mathcal{X})$. These quantities are local functions of $\mathcal{X}$, whose expectation value can be calculated exactly given $Z[\mathcal{H}]$. The real-space RG procedure thus performed is therefore an exact technique, which, in particular, always preserves critical behavior, regardless of what $\Lambda$ is. 

Notwithstanding, the usefulness (and practicality) of the RG procedure depends on choosing $P_{\Lambda}(h_i | \mathcal{X})$ (or equivalently the relevant degrees of freedom) such that effective Hamiltonian ${\rm \tilde{H}}$ remains as short range as possible and (if $\mathcal{H}$ is continuous) the fluctuations of $\mathcal{H}$ are as small as possible, so that high powers of $h_i$ are not needed. More formally we demand that the Taylor expansion of ${\rm \tilde{H}}$ in $h_i$:
\begin{equation}
{\rm \tilde{H}} = \sum_i {\rm f}_{i}h_i + \sum_{\langle ij \rangle} {\rm f}_{ij}h_ih_j + \sum_{\langle \langle ij \rangle \rangle } {\rm f}_{ij}h_ih_j
+\sum_{\langle ijk \rangle}{\rm f}_{ijk}h_ih_jh_k + \sum_{\langle \langle ijk \rangle \rangle} ...  
\label{Eq:taylor1}
\end{equation}
is as compact as possible i..e. it contains only short-ranged and few-body terms (the coefficients $f$ decay exponentially with distance).  If the above requirements are satisfied, all the terms in ${\rm \tilde{H}}$ beyond some finite distance can be removed while making only minor changes to statistical properties of  system $\mathcal{H}$. The procedure can then be repeated recursively granting access to increasingly long-ranged features without keeping track of all degrees of freedom.

What constitutes a good RG scheme in the above language is intuitively clear, but hard to formalise, especially with
an algorithmic goal in mind. 
The mutual information maximization (MIM) prescription, on the other hand, has a very precise formulation, and lends itself naturally to computational implementation. We argue that the two approaches are equivalent. Intuitively it is because maximizing the mutual information encourages $h_i$ to couple to the combinations of $\mathcal{V}$ which are most strongly correlated with the environment. In field theory terms these combinations are the most relevant operators in the theory which are nothing else than the basic fields appearing in the low energy theory. Clearly, when performing an RG scheme one needs to keep track of precisely those fields.  The numerical results reported in the main text provide strong empirical evidence for this equivalence. Below we provide two additional analytical insights which reinforce this assertion: (1) For the one-dimensional Ising model our approach coincides with the standard ``decimation" RG approach, which is known to be optimal (2) For 1D (and quasi-1D) systems with nearest-neighbour interactions saturating the mutual information implies an effective Hamiltonian with nearest-neighbor interactions.

Lastly, we point out that mutual information is invariant under any homeomorphism of the degrees of freedom in $\mathcal{X}$ or $\mathcal{E}$. Consequently, scrambling the degrees of freedom in $\mathcal{X}$, or taking complicated non-linear functions of them, does  not affect this procedure (i.e. it is representation-invariant, as it should be). More precisely, if $P(h|\mathcal{X})$ maximizes the mutual information w.r.t. $\mathcal{X}$ and $P'(h|\mathcal{X'})$ maximizes the mutual information w.r.t. $\mathcal{X'} = f(\mathcal{X})$ then $P(h|\mathcal{X}) = P'(h|f^{-1}(\mathcal{X'}))$.

\subsection*{1. Ideal filters for the 1D Ising model}
The 1D Ising model is one of the simplest statistical mechanical models; it can be solved exactly. In particular, the decimation RG prescription is known to be optimal for this model.
We shall determine what the ideal filters -- in terms of mutual information -- are, and show that they indeed correspond to decimation.  

The partition function on $L$ sites (periodic boundary conditions) is given in the transfer matrix formalism by:
\begin{equation}
Z = Tr[T^N],  \ \ \ \ \ 
T = \left(\begin{array}{cc}
e^{K} & e^{-K}  \\
e^{-K} & e^{K} \\
\end{array}\right) \mbox{\ \ \ with:}
\end{equation} 
\begin{equation}
T^m = [\cosh^{2m}(K)-\sinh^{2m}(K)]^{1/2}\left(\begin{array}{cc}
\left(\frac{\cosh^m(K) + \sinh^m(K)}{\cosh^m(K) - \sinh^m(K)}\right)^{1/2} & \left(\frac{\cosh^m(K) + \sinh^m(K)}{\cosh^m(K) - \sinh^m(K)}\right)^{-1/2}   \\
\left(\frac{\cosh^m(K) + \sinh^m(K)}{\cosh^m(K) - \sinh^m(K)}\right)^{-1/2} &\left(\frac{\cosh^m(K) + \sinh^m(K)}{\cosh^m(K) - \sinh^m(K)}\right)^{1/2} \\
\end{array}\right).
\end{equation} 

The dominant eigenvector of $T$ is $|g\rangle = (1,1)$ with an eigenvalue $2 \cosh(K)$. Note that $T^m$ is proportional to $T$ with an effective $K_m = \frac{1}{2} \log \left(\frac{\cosh^m(K) + \sinh^m(K)}{\cosh^m(K) - \sinh^m(K)}\right)$ which always tends to zero at large $m$. This implies that the model is always short range correlated.

Let us  now consider the following geometry:  a line region $\mathcal{V}=\{v_1...v_{L_{\mathcal{V}}}\}$ of size $L_{\mathcal{V}}$ consisting of spins $v_i \in \{-1,1\}$, with two buffer zones  of size $L_B$ to its left and right, surrounded by an environment  $\mathcal{E}=\{e_{-L_{\mathcal{E}}},...,e_{-1},e_{1},...,e_{L_{\mathcal{E}}} \}$ of size $2 L_{\mathcal{E}}$ and then by the rest of the system which is assumed to be much larger than the correlation length. The probabilities involved in maximizing the mutual information can now be written down explicitly :
\begin{equation}
P_{\Theta}(\mathcal{V}) = \frac{\langle g | T |v_1\rangle T_{v_1,v_2}T_{v_2,v_3}...T_{v_{L_{\mathcal{V}}-1},v_{L_{\mathcal{V}}}} \langle v_{L_{\mathcal{V}}} | T | g \rangle}{\left[ 2 \cosh(K) \right]^{L_{\mathcal{V}}+1}},
\end{equation}
\begin{equation}
P_{\Theta}({\mathcal{V}},{\mathcal{E}}) = \frac{\langle g | T | e_{-L_{\mathcal{E}}} \rangle T_{e_{-L_{\mathcal{E}}},e_{-L_{\mathcal{E}}+1}}...T^{L_B+1}_{e_{-1},v_1}T_{v_1,v_2}...T^{L_B+1}_{v_{L_{\mathcal{V}}},e_1}..T_{e_{L_{\mathcal{E}}-1},e_{L_{\mathcal{E}}}} \langle e_{L_{\mathcal{E}}} T | g \rangle}{\left[ 2 \cosh(K) \right]^{L_{\mathcal{V}}+2L_B+2L_{\mathcal{E}}+1}},
\end{equation}
Using the definitions of $T$ and $|g\rangle$  we find:  
\begin{eqnarray}
{\rm E_{\Theta}}(\mathcal{V}) &=& -\sum_{v_i \in \mathcal{V}} K v_i v_{i+1}  + {\it const}, \\
{\rm E_{\Theta}}({\mathcal{E}},\mathcal{V}) &=& -\sum_{e_i \in {\mathcal{E}}} K e_i e_{i+1} - K_{L_B} \left(e_{-1} v_1 + v_{L_{\mathcal{V}}} e_1\right) + {\rm E_{\Theta}}(\mathcal{V}) +  {\it const'},\\
\Delta {\rm E_{\Theta}} &=& -\sum_{e_i \in {\mathcal{E}}} K e_i e_{i+1} -K_{L_B} \left(e_{-1} v_1 + v_{L_{\mathcal{V}}} e_1\right) + {\it const''},
\end{eqnarray}
where the energies are defined via  $P \propto e^{-E}$. We now have everything needed to evaluate the MI proxy $A_{\Lambda}$:
\begin{equation} A_{\Lambda}(\mathcal{H}:\mathcal{E}) \approx \sum_{\mathcal{H},\mathcal{V},\mathcal{E}} P_{\Theta}(\mathcal{E},\mathcal{V})P_{\Lambda}(\mathcal{H}|\mathcal{V}) \left\langle - \Delta E_{\Theta}(\mathcal{V},\mathcal{E})\right\rangle_{\mathcal{H}}. \end{equation}

Let us examine this expression: any term in $\Delta E_{\Theta}$ which does not involve $\mathcal{V}$, can be taken out of the average. For instance $\langle K e_{-2} e_{-1} \rangle_{\mathcal{H}} = K e_{-2} e_{-1} $. Furthermore, the summation $\sum_{\mathcal{H}}P(\mathcal{H}|\mathcal{V})$ can then be performed trivially, yielding a factor of $1$. Thus the $\mathcal{V}$-independent terms are also $\Lambda$-independent and therefore they can be ignored as they are irrelevant for minimization w.r.t. $\Lambda$. We may therefore redefine $\Delta E_{\Theta}$ as:
\begin{equation}
\Delta {\rm E_{\Theta}} = -K_{L_B} \left(e_{-1} v_1 - v_{L_{\mathcal{V}}} e_1\right). 
\end{equation}
Similar arguments imply that one may redefine ${\rm E_{\Theta}}({\mathcal{E}},\mathcal{V})$ as: 
\begin{equation}
{\rm E_{\Theta}}({\mathcal{E}},\mathcal{V}) =  -K_{L_B} \left(e_{-1} v_1 - v_{L_{\mathcal{V}}} e_1\right) + {\rm E_{\Theta}}(\mathcal{V}) 
\end{equation}

The exact solution to the problem of maximization of $A_{\Lambda}(\mathcal{H}:\mathcal{E})$ is difficult to find analytically,  even for the case of a single hidden variable $h$. We  may, however, evaluate and compare the three most likely scenarios: 

{\bf Decimation filter:} It is defined by coupling to a single visible spin only. Here we assume, without loss of generality, that $P(h|v_1,v_{L_{\mathcal{V}}})=1$ iff $h=v_1$. Consequently: 
\begin{equation}
\langle -\Delta E \rangle_{\mathcal{H}} = K_{L_B} e_{-1} h + K_{L_B} e_1 \langle v_{L_{\mathcal{V}}} \rangle_{{\rm E_{\Theta}}(\mathcal{V}),v_1=h}
\end{equation}
where the last average is taken with respect to $e^{-{\rm E_{\Theta}}(\mathcal{V})}$ with the constraint $v_1=h$. For large $L_{\mathcal{V}}$ this last term would be exponentially small in $L_{\mathcal{V}}$ and can be neglected.  We therefore obtain:
\begin{equation}
A_{\Lambda}(\mathcal{H}:\mathcal{E}) \approx \sum_{\mathcal{V},\mathcal{E}} P(\mathcal{V},\mathcal{E}) K_{L_B} e_{-1} v_1 = K_{L_B} \langle e_{-1} v_1 \rangle,
\end{equation}
where the last average is taken with respect to the partition function of the 1D Ising model. 

{\bf Boundary filter:}. It is defined by equal coupling to the two boundary spins, i.e. the coarse grained variable $h$ is determined by majority rule from the $v_i$ on the boundary:
\begin{equation}
P(h|v_1,v_{L_{\mathcal{V}}}) =  \frac{1 + h(v_1+v_{L_{\mathcal{V}}})}{2}.
\end{equation}
First, we evaluate the quantity: 
\begin{equation}
\langle -{\rm \Delta E}_{\Theta} \rangle_{\mathcal{H}} = K_{L_B} \left ( e_{-1} \langle v_1 \rangle_{\mathcal{H}} +  e_{1} \langle v_{L_{\mathcal{V}}} \rangle_{\mathcal{H}} \right).
\end{equation}
Neglecting the correlations between $v_1$ and $v_{L_{\mathcal{V}}}$ vanishing for large $L_{\mathcal{V}}$ we have with $P(v_1,v_{L_\mathcal{V}})=1/4$. Consequently $\langle v_1 \rangle_{\mathcal{H}} = \langle v_{L_{\mathcal{V}}} \rangle_{\mathcal{H}} = h/4$ and we obtain: 
\begin{equation}
\langle {\rm \Delta E}_{\Theta} \rangle_{\mathcal{H}} = K_{L_B} \left ( e_{-1} + e_{1} \right) \frac{h}{4}.
\end{equation}  
Finally, we have for $A_{\Lambda}$:
\begin{equation}
A_{\Lambda}(\mathcal{H}:\mathcal{E}) = 2 \frac{K_{L_B}}{4} \sum_{\mathcal{V},h,\mathcal{E}} P(h|v_1,v_{L_{\mathcal{V}}}) P(\mathcal{V},\mathcal{E}) e_{-1} h = 2 \frac{K_{L_B}}{4} \sum_{\mathcal{V},h,\mathcal{E}}  \frac{1 + h(v_1+v_{L_{\mathcal{V}}})}{2} P(\mathcal{V},\mathcal{E}) e_{-1} h,
\end{equation}  
yielding:
\begin{equation}
A_{\Lambda}(\mathcal{H}:\mathcal{E}) = \frac{K_{L_B}}{2} \langle e_{-1}v_1 \rangle \end{equation} 

{\bf Uniform filter:}. It is defined by uniform coupling to all visible spins, or equivalently, $h$ is determined  by majority rule on all of $v_i$:
\begin{equation}
P(h|\mathcal{V}) =  \frac{1 + h\, \text{arctanh} (\Omega \sum_i v_i)}{2}
\end{equation}
with $\Omega \gg 1$.   
We again need to evaluate the quantity:
\begin{equation}
\langle -{\rm \Delta E}_{\Theta} \rangle_{\mathcal{H}} = K_{L_B} \left ( e_{-1} \langle v_1 \rangle_{\mathcal{H}} +  e_{1} \langle v_{L_{\mathcal{V}}} \rangle_{\mathcal{H}} \right),
\end{equation}
where the averages on the r.h.s. are now taken using $P(\mathcal{V})$ but constrained to have the majority of $v_i$ aligned with $h$. To simplify computations we exchange this hard constraint for an equivalent  soft one: to this end we introduce a fictitious magnetic field $H=f(h)$ and demand that it generates the same average magnetization as $h$ does via the hard constraint. As such approximations are only valid in the thermodynamical limit, the important quantity to track here is the scaling of all quantities with $L_{\mathcal{V}}$. The hard constraint induces an average magnetization proportional to the square-root of the variance of the magnetization in the absence of the constraint and so is proportional to $\sqrt{L_{\mathcal{V}}}$.  On the other hand $H$ induces magnetization proportional to $L_{\mathcal{V}}$, as it couples to all spins directly. As a result, we need $H \propto h/\sqrt{L_{\mathcal{V}}}$ to reproduce the hard constraint,  and we can then estimate:
\begin{equation}
\langle -{\rm \Delta E}_{\Theta} \rangle_{\mathcal{H}} \propto  \frac{K_{L_B}}{\sqrt{L_{\mathcal{V}}}} e_{-1} h ,
\end{equation}
and  thus: 
\begin{equation}
A_{\Lambda}(\mathcal{H}:\mathcal{E}) \propto \frac{K_{L_B}}{\sqrt{L_{\mathcal{V}}}} \sum_{\mathcal{V},h,\mathcal{E}} P(h|v_1,v_{L_{\mathcal{V}}}) P(\mathcal{V},\mathcal{E}) e_{-1} h .
\end{equation}  
Since the quantities being averaged are bounded by a constant, $A_{\Lambda}$ vanishes in the limit of large $L_{\mathcal{V}}$. 

Comparing the results in the three cases we conclude that the decimation filters are favored,  as they yield twice the mutual information compared to the boundary filters (both are superior to the uniform filter). These results easily generalize to the anti-ferromagentic Ising model. We also confirmed the results by exact numerical evaluation for small systems. Interestingly, the decimation filters are known to be optimal from analytical point of view as they result  in an effective Hamiltonian with nearest-neighbour interactions only. Hence, the mutual information RG scheme coincides with the standard result.

\subsection*{2. MI saturation implies a nearest-neighbour Hamiltonian} 

Consider a 1D, or quasi-1D, system $\mathcal{X}$, with a spatial subset of degrees of freedom $\mathcal{V}$ separating two parts of the environment $\mathcal{E}_1$ and $\mathcal{E}_2$, and the hiddens $\mathcal{H}$ coupled to $\mathcal{V}$ via parameters $\Lambda$  defined by the MIM prescription. Let us assume that adding further hiddens does not cause the mutual information to grow, i.e. MI has saturated.   We argue that if this is the case, the effective Hamiltonian contains no interaction between any $h_i$ and $h_j$ which couple exclusively to distinct parts of the environment,  $\mathcal{E}_1$ and $\mathcal{E}_2$, respectively. 

To this end consider the conditional probability distribution $P(\mathcal{H} | \ \mathcal{V}|_{r_n})$, where $\mathcal{V}|_{r_n}$ is a box-shaped area to be coarse-grained, centered around site $r_n$ (similar to the one depicted in Fig. \ref{fig:multisteps}):
\begin{equation}
P(\mathcal{H} | \ \mathcal{V}|_{r_n}) = \frac{e^{\sum \lambda_i h_i(r_n) O_i(\mathcal{V}|_{r_n})}}{\Pi_i 2 \cosh \left[\lambda_i O_i(\mathcal{V}|_{r_n})\right]} ,
\end{equation} 
and where $O_i(\mathcal{V}|_{r_n})$ are  functions of visibles in that area. The coefficients ${\rm f}_{ij..}$  in the effective Hamiltonian Eq. \ref{Eq:taylor1} now appear as a cumulants with respect to ${\rm H}_{massive}$ given by:
\begin{equation}
{\rm H}_{massive} = {\rm H} - \sum_{n} \sum_i \log \cosh(\lambda_i O_i(\mathcal{V}|_{r_n})).
\end{equation} 
These additional terms can be expanded assuming small $O_i(\mathcal{V})$ fluctuations, yielding: 
\begin{equation}
{\rm H}_{massive} \approx {\rm H} + \sum_{n,i} (\lambda_i O_i(\mathcal{V}|_{r_n}))^2,
\end{equation}
whilst for stronger fluctuations the mass term has a different asymptotic $|\lambda_i O_i (\mathcal{V}_{r_n})|$. 
Crucially, both of these have the property that when $\lambda_i$ tends to infinity, the fluctuations of $O_i(\mathcal{V}|_{r_n})$ are completely suppresed. The interaction in the effective Hamiltonian ${\rm \tilde{H}}$ between two hiddens  $h_i(r_n)$ and $h_j(r_m)$ coupling to visible areas centered around $r_n$, $r_m$  can now be expressed as: 
\begin{equation}
{\rm f}_{(ni),(mi)} = \langle O_i(\mathcal{V}|_{r_n}) O_j(\mathcal{V}|_{r_m}) \rangle_{{\rm H}_{massive},connected}
\label{Eq:connected1}
\end{equation}
where the r.h.s. is a connected correlation function (i.e. with all averages properly removed) under the Hamiltonian ${\rm H}_{massive}$. 

Let us now assume that  $r_n$ is within the environment $\mathcal{E}_1$ and $r_m$ is within $\mathcal{E}_2$ and nevertheless ${\rm f}_{(ni),(mj)} \not= 0$.  Eq. \ref{Eq:connected1} implies that some correlation remains between $O_i(\mathcal{E}_1)$ and $O_j(\mathcal{E}_2)$. However, in a local statistical mechanical system, with strictly short range interactions, correlations cannot skip over a region. Consequently, there exists an operator $O_*$ with support in $\mathcal{V}$, which is also correlated with $\mathcal{E}_1$ and $\mathcal{E}_2$ {\it even under the massive Hamiltonian} ${\rm H}_{massive}$ which has all $O_i$ essentially frozen to zero. If $O_* = \sum_i a_i O_i(\mathcal{V})$, then its fluctuations should have been suppressed by the existing mass terms since MI has been maximized. Alternatively if it is not a linear combination of the previous operators then clearly there is mutual information to be gained by adding an additional hidden degree $h_*$ which couples to $O_*$. This, however, contradicts the assumption of MI saturation. We therefore conclude that upon saturating the mutual information no next-nearest-neighbor interactions can appear in the effective Hamiltonian.

\section*{A comment on uniform vs. boundary-centered filters for 2D Ising model}

In the previous section we found that ``decimation'' filters are optimal for the 1D Ising model from the point of view of mutual information. For the 2D models we studied, we also numerically compared the behaviour of the weights $\lambda_i^j$ for areas $\mathcal{V}$ of increasing size, which are generalizations of Kadanoff procedure to larger blocks. We find the network couples to the boundaries of the area $\mathcal{V}$ to maximize MI with the rest of the system [see Figs. \ref{fig:ising2}(C,D,E)]. This physical insight the RSMI provides can in fact be shown to hold exactly for a standard real-space RG in the limit of number of coarse grained variables equaling the size of the boundary (see below). 

\begin{figure}[h]
	\centering
	\includegraphics*[width=12cm]{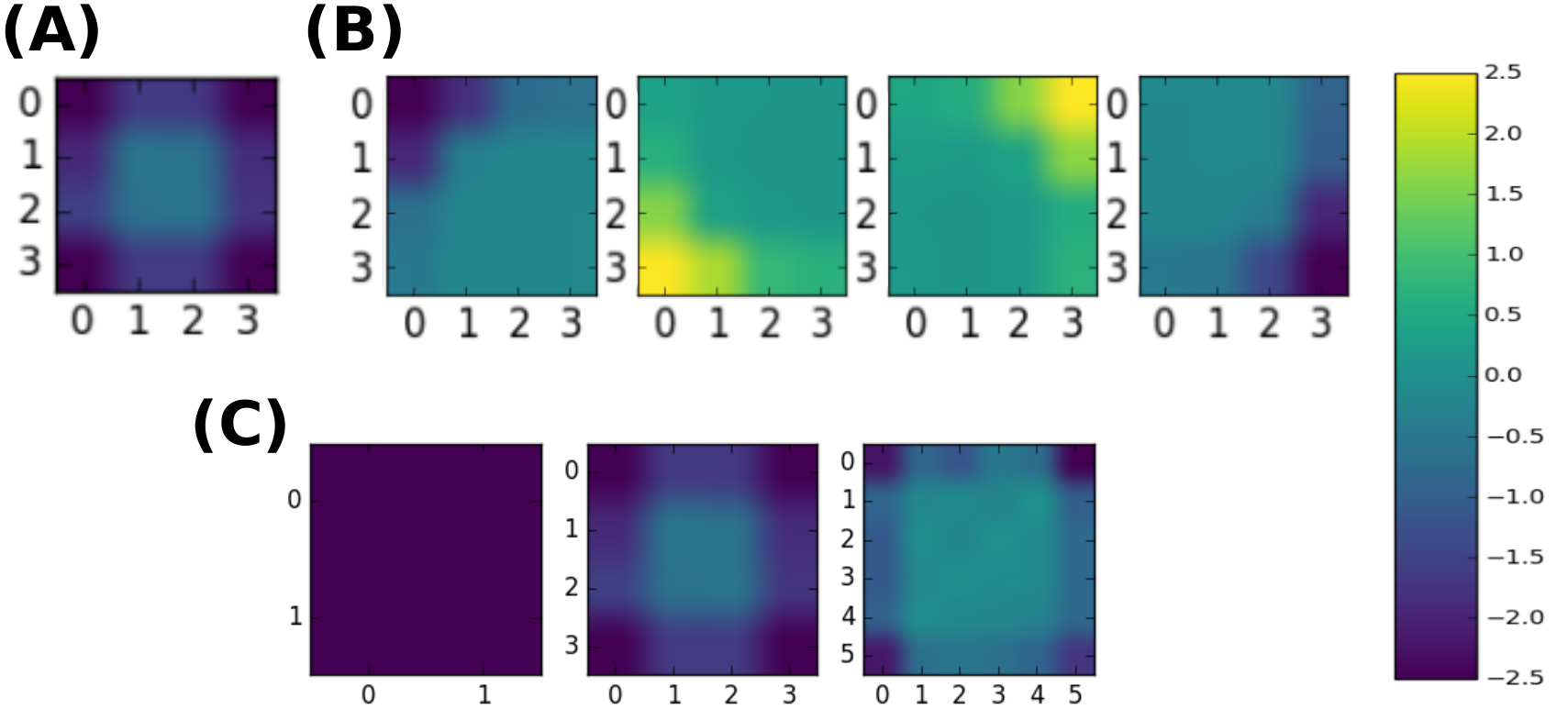}
	\caption{The weights of the RSMI network trained on Ising model data. The network couples strongly to areas with  large absolute value of the weights. (A) $N_h=1$ for a $4 \times 4$ visible area. (B) $N_h=4$ for a $4 \times 4$ visible area. (C) Comparison of $N_h=1$ weights for area size of $2 \times 2$, $4 \times 4$, $6 \times 6$ -- the network couples to the boundary. } 
	\label{fig:ising2}
\end{figure}

Let us investigate the following question: should the parameters $\lambda_i^j$  defining slow degrees of freedom in a real-space RG scheme [as described in the previous section of Supplemental Materials around Eqs. (12,13)], in the limit of a large area $\mathcal{V}$, have a uniform modulation (pattern) over the area $\mathcal{V}$ or should they more strongly couple to the boundary of $\mathcal{V}$? We argue that for the 2D Ising case it is the latter, therefore the results obtained by our MI-based procedure are consistent with behaviour expected of an RG scheme.
For the sake of argument we assume the simplest scenario, when the system has only short range interactions and the number of ``hiddens" coupling to $\mathcal{V}$ equal the number of degrees of freedom on its boundary $\partial\mathcal{V}$ (defined as the set of degrees of freedom in $\mathcal{V}$ with neighbours outside $\mathcal{V}$). It is easy to see that in this case the optimal solution is to couple  each hidden unit $h_{i}$,  with infinite strength, to one degree of freedom $v_{i}$  in $\partial\mathcal{V}$, effectively enforcing $h_i \equiv v_{i}$ there. In the spirit of the discussion in the previous section of Supplemental Materials, integrating out the $h_i$ would generate large mass terms for the $v_i$ on the boundary, removing all correlations between the inside and outside of $\mathcal{V}$ and consequently the effective Hamiltonian for the $h_i$ would be short-ranged. For number of hidden units smaller than $\partial{V}$, the filters would keep favoring the boundary over the bulk of the cell to keep track of the degrees of freedom most directly coupled to the environment, an intuition which is qualitatively confirmed by our numerical results for the Ising model.  

For the case of the dimer system the weight matrices appear uniformely textured even for a large area $\mathcal{V}$ [see Fig. 4(A) in the main text], which may seem puzzling in light of the above argument. This is the result of dimers being constrained degrees of freedom (or, in the continuum formulation, the height field obeying a conservation equation) \cite{Fradkin}, as opposed to Ising spins. The height field $h$ is not a microscopically accessible quantity in dimer model, only its gradient $\nabla h$ is. The weights $\lambda_i^j$ we find for the dimer model enforce a uniform coupling pattern of the hidden degrees of freedom to the gradient of height field over many unit cells contained in the visible area $\mathcal{V}$, i.e. the hiddens couple to an average of the electric field in $\mathcal{V}$. This integral of $\nabla h$ over $\mathcal{V}$ is, however, equivalent to a difference of boundary terms for the height field $h$ itself! Thus there is no contradiction: the direct coupling of hiddens to the height field on the boundary, which is physically not possible for the dimer model, is achieved by a uniform coupling to its gradient. 

In principle, the type of scaling behaviour under the size of $\mathcal{V}$ may allow to draw conclusions as to whether the degrees of freedom are constrained or not.

\section*{Comparison with Contrastive divergence trained RBMs}

The relation between unsupervised machine learning (in particular in deep ANNs), and RG, has been a subject of some controversy recently. Ref.\cite{mehta} claims  an ``exact'' mapping between the two, while Tegmark and Lin assert, to the contrary,  that RG is entirely unrelated to unsupervised learning \cite{tegmark}. Our theoretical arguments, as well as explicit numerical results on dimer model disprove the very general claims of Ref. \cite{mehta}, however, we cannot entirely agree with Tegmark and Lin either.  
\begin{figure}[h]
	\centering
	\includegraphics*[width=9cm]{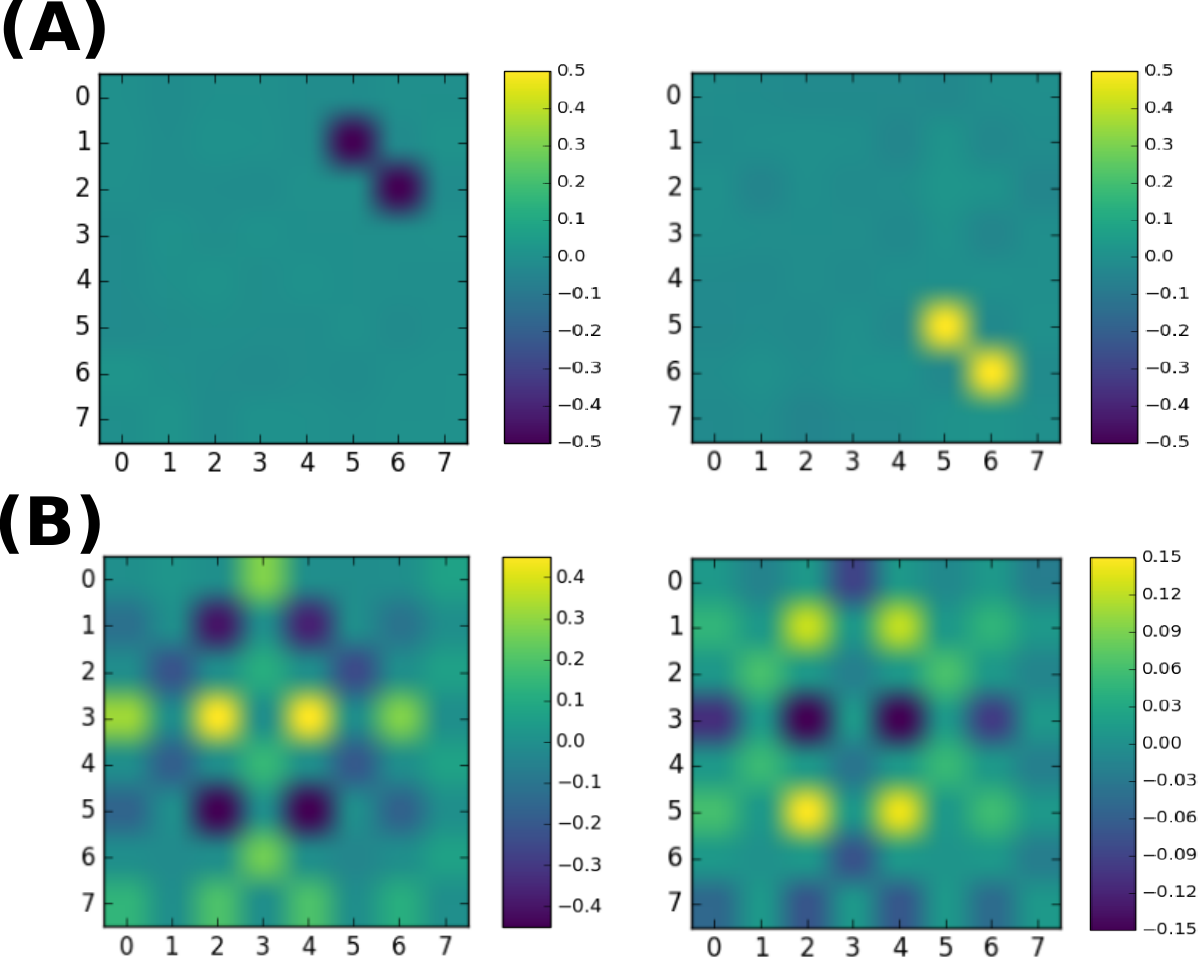}
	\caption{The weights of a standard CD-trained RBM on dimer model with additional noise: (A) For a small number of hidden neurons the network learns to recognize the spin-pair pattern (B) Given enough filters to capture all the spin pairs, the successive filters capture dimer configurations more similar to the ``columnar'' state in right pannel of Fig. 3(B), instead of the staggered configurations the RSMI finds.}  
	\label{fig:ising3}
\end{figure}

The crucial point is that a neural network is not specified without  the cost function.  Both Ref. \cite{mehta} (explicitly) and Ref. \cite{tegmark} (implicitly) assume that the networks are trained using Kullback-Leibler divergence, or some related measure, resulting in the network which approximates well the input data probability distribution. Ref. \cite{tegmark} is correct in pointing out that this will not generally result in RG transformation. We are not limited, however, to such cost functions; in point of fact, we are free to chose a cost function with an entirely different objective. As we have shown, chosing to maximize the mutual information with the system at large results in a network identifying the physically relevant degrees of freedom.

To disprove explicitly that a generic NN trained to recognize patterns performs a physically meaningful coarse-graining (i.e., potentially, RG), we examine the weight matrices of CD-trained network ($\Theta$-RBM) on the dimer model with additional spin ``noise", the same we considered in the main text. In Fig. 6(A) we show the examples of weights obtained for a small number of hidden units: the network strongly couples to individual pairs of spins fluctuating in sync, even though they are irrelevant from the point of view of physics. This is correct behaviour, when patterns are to be discerened (since there are many entirely different dimer textures, but the same fluctuating pattern of spin pairs for each configuration), but does not make any sense physically. If given a few hidden units we were to extract new degrees of freedom to further continue the RG procedure on, we would discard \emph{all of} the dimer model in the first step, ending up with a trivial system of decoupled spin pairs. Only given enough hidden neurons to capture all the spin pairs does the CD-trained network learn to discern extensive dimer textures, and even then they are not the staggered configurations giving electric fields important from the point of RG, but rather more similar to the high-entropy columnar configurations (which map to vanishing electric fields).

\end{document}